\documentclass{article}

\usepackage{PRIMEarxiv}

\usepackage[utf8]{inputenc} 
\usepackage[T1]{fontenc}    
\usepackage{hyperref}       
\usepackage{url}            
\usepackage{booktabs}       
\usepackage{amsfonts}       
\usepackage{nicefrac}       
\usepackage{microtype}      
\usepackage{lipsum}
\usepackage{fancyhdr}       
\usepackage{graphicx}       
\graphicspath{{media/}}     

\usepackage{makecell}
\usepackage{multirow}
\usepackage{tabularx}
\usepackage{threeparttable}
\usepackage{float}
\usepackage{placeins}
\usepackage{tabularx,booktabs}
\usepackage{adjustbox}

\title{Physical Unclonable Functions (PUF) for IoT Devices
}

\author{
  Abdulaziz Al-Meer, Saif Al-Kuwari \\
  Division of Information and Computing Technology, College of Science and Engineering \\
  Hamad Bin Khalifa University, Qatar Foundation,   Doha-Qatar \\
  abal34183@hbku.edu.qa, smalkuwari@hbku.edu.qa \\
}

\begin{document}
\maketitle

\begin{abstract}
Physical Unclonable Function (PUF) has recently attracted interested from both industry and academia as a potential alternative approach to secure Internet of Things (IoT) devices from the more traditional computational based approach using conventional cryptography. PUF is promising solution for lightweight security, where the manufacturing fluctuation process of IC is used to improve the security of IoT as it provides low complexity design and preserves secrecy. It provides less cost of computational resources which prevent high power consumption and can be implemented in both Field Programmable Gate Arrays (FPGA) and Application-Specific Integrated Circuits (ASICs). In this survey we provide a comprehensive review of the state-of-the-art of PUF, its architectures, protocols and security for IoT. 
\end{abstract}


\section{Introduction}\label{Introduction}

Preserving authenticity and confidentiality in IoT networks  are major concerns for researchers and industries due to the increase reliance on these networks at many applications such as, sensors, monitoring and healthcare \cite{D.wang:2019}.

To address these concerns, PUF approaches have recently been proposed as a promising alternative to conventional cryptography, which is largely  unsuitable to IoT network mainly because of its sheer computational requirement, which often degrade battery efficiency on the lightweight IoT devices. 

Additionally, conventional cryptography is based on the hardness of solving some mathematical problems. However, advances in quantum computers may render such problems easy to solve \cite{J.zhang:2019}\cite{mahdi:2020}\cite{Shor-P.W:1994}.

Moreover, key distribution techniques used in conventional cryptography usually require a third party, which makes it impractical for IoT applications \cite{N.xie:2021} as large-scale wireless network require keys to be frequently updated, and that will introduce high overhead on the network. 

Finally, some conventional cryptography primitives such as, message authentication code (MAC), rely on the upper-layer mechanisms, which can be manipulated by adversaries e.g., spoofing attack.

On the other hand, PUF offers an alliterative solution for lightweight hardware security suitable for IoT networks. PUF is based on the fact that no two identical chips having the same characterization and going through the same production line will never share the same physical properties to inevitable manufacturing variation during the fabrication of the chips. \cite{Schinianakis:2019}. PUF circuits exploit these features and utilize such variations to generate secret keys and authenticate IoT devices.

The rest of the survey is organized as following: section \ref{Fundamental} provides a brief discussion on the fundamentals of PUF. Section \ref{PUF performance} describes PUF performance evaluation and quality metrics. PUF authentication protocols are then introduced in section \ref{PUF authentication} followed by a thorough description of the popular architectures of PUF in section \ref{PUF Architectures} with analysis of specific security requirements such as, strengths and weakness for each circuit design as well as performance evaluation comparison. In section \ref{PUF implementation}, we discuss common PUF implementations, mainly focusing on FPGA implementation. Section \ref{Threat landscape} discuss threats and common attack against PUF. Finally, we conclude this paper in section \ref{Conclusion} and discuss some potential future research directions.

\subsection{Related work} 

Several survey were published on PUF considering both current PUF devices and emerging technologies. Babaei and Schiele \cite{babaei:2019} presented an overview of PUF for authenticating IoT devices and investigated the related challenges toward PUF exploitation. Similarly, the survey in \cite{zhang_qu:2014} highlighted various silicon PUF, mainly on Ring Oscillator (RO) PUFs with related issues and challenges. Furthermore, Chang \emph{et al.} \cite{Chang:2017} reviewed the improvement of PUF over past decade and demonstrate vulnerabilities of PUF. Halak \emph{et al.} \cite{Halak:2016} presented an overview of PUF in term of principle and design challenges. A tutorial on PUF applications, error correction mechanism, PUF types and emerging technologies were presented in \cite{Herder:2014}. Moreover, the authors in \cite{Shamsoshoara:2020} presented a review of the IoT network security challenges and investigate related attacks based on several IoT domains and discussed fuzzy extractors schemes for key extractions. In addition, Alkatheiri \emph{et al.} \cite{Alkatheiri:2017} presented an experimental study of three designs in each of the two categories of PUFs: delay based and frequency variation PUFs. Table \ref{tab:comparison} provides a comparison between previous survey and ours.

\subsection{Motivation}

One of the attractive lightweight security solutions for IoT devices is PUF. Consequently, many researchers discussed and reviewed the emerging PUF technologies and their security challenges. 

In addition, several PUF architectures have been proposed in the recent years. 
However, to the best to our knowledge, there is no comprehensive review of PUF that discusses important aspects of PUF, such as the recent PUF implementations, quality evaluations and security perspective for common and recent attack for different PUF architectures. This has motivated us to write this survey and provide such recent review for this rather important and emerging technology.

\subsection{Our contributions}

Our contributions of this survey are as following:
\begin{itemize} 
   \item We investigate the essential performance evaluation and quality metrics of different PUF category and design.
   \item We introduce recent PUF authentication protocols and compare them for lightweight devices, while showing how they mitigate common attack.
   \item We discuss different PUF architectures suitable for IoT applications, specially PUF designs implemented on FPGAs, which is becoming an attractive development platform.
   \item We investigate the most common threats and attack on PUF and discussed multiple assumptions and scenario.
   \item Finally, we discuss some open problems, identity gaps and make recommendations and directions for future work. 
\end{itemize}

\begin{table*}[ht]
  \caption{Previous Survey Comparison and contributions}
  \label{tab:comparison}
  \resizebox{\textwidth}{!} {%
  \begin{tabular}{ccccccc}
    \toprule
    Ref. & PUF Applications  & PUF evaluation  & PUF protocol & PUF Architectures & PUF implementation & PUF Threats and attack \\
    \midrule
    \makecell{\cite{Herder:2014}} & \checkmark & x & \checkmark  & \checkmark & x & \checkmark   \\ 
    \makecell{\cite{zhang_qu:2014}} & \checkmark  & x & \checkmark & \checkmark & x & \checkmark  \\ 
    \makecell{\cite{Adames:2016}} & x & \checkmark & x & \checkmark & \checkmark & x \\ 
    \makecell{\cite{Halak:2016}} & \checkmark & x & \checkmark & x & x & x \\ 
    \makecell{\cite{Chang:2017}} & \checkmark & x & \checkmark & \checkmark & x & \checkmark \\ 
    \makecell{\cite{Alkatheiri:2017}} & x & \checkmark & x & \checkmark & \checkmark & x \\ 
    \makecell{\cite{babaei:2019}}  & \checkmark & x & \checkmark  & \checkmark & x  & \checkmark  \\ 
    \makecell{\cite{Shamsoshoara:2020}} & \checkmark & x & x & \checkmark & x & \checkmark  \\ 
    \makecell{This work} & \checkmark  & \checkmark & \checkmark & \checkmark & \checkmark & \checkmark  \\ 
    \bottomrule
  \end{tabular}%
  }
\end{table*}

\section{PUF fundamentals}\label{Fundamental}

\subsection{Definitions}\label{Definitions}
The inherent unclonability of the PUF cannot be controlled as it is based on multiple random parameters that are generated during the manufacturing process. When the binary sequence applied to PUF system, it will react with corresponding response. That is, no two Integrated Circuits (IC) provide an identical response $R$ for the same challenge $C$ and this combination is called challenge-response pair (CRP). The PUF system contains uncontrollable random components, so when the challenge $C$ applied to the PUF system it will react with these components in a way to produce unpredictable and random response $R$. 

These random components and the inability to control the manufacturing process make the PUF system unpredictable, unique, and more important, Physically Unclonable\cite{Chang:2016}.

PUF system is considered physically disordered with the structural information as following:
\begin{itemize} 
   \item The related information of the PUF system could be extracted in a reliable way through measurement when different challenge $C$ applied to the system to generate identical response $R$.
   \item Due to the large possible combination of challenges $C$, the corresponding response $R$ cannot be predictable within restricted time.
   \item It is very hard and almost impossible to model, computationally and numerically determine and predict the challenge-response pair (CRP) based on the available information and current pairs.
   \item The PUF system cannot be cloned and reproduced due to the variation of the manufacturing process. 
\end{itemize}

The physical characteristics of the PUF circuits can vary in term of: signals transmission speed, frequency oscillation, and the initial random state of the memory elements. These characteristics can be exploited to Physical Layer Security (PLS).

\subsection{PUF Applications}
The main goal of PUF is to ensure communication security and prevent possible attack. There are several applications of PUF that can be utilized for identification, confidentiality and to authenticate. Below we list some popular PUF application, each of which may prefer some PUF designs over others based on their requirements: 
\begin{itemize} 
   \item \emph{True Random number sequence generator}: usually used to generate keys for encryption in communication and digital signatures and create password to protect the system\cite{Suh:2007}.
   \item \emph{Malware detection}: Malware detection is one of the most time-consuming processes in hardware security. However, PUF can efficiently detect differences between original devices and malware injected device by examining the corresponding challenge-response pairs \cite{Tehranipoor:2010}. That is, when a chip is injected with malware, it will inevitably changes the power distribution of the device and that will deviate the response of the chip.
   \item \emph{Detection of degraded hardware performance}: The performance of the chip can be degraded due to natural aging and time, as a result PUF challenge-response pair can detect such devices specially with the critical applications such as, aviation, military, and healthcare.
   \item \emph{Hand weapon authentication}: Utilizing embedded PUF device to authenticate, secure the authorized users of a weapon. For instance, if the weapons are lost, no one other than its original owner will be able to use it \cite{armatix:2021}.
   \item \emph{Self-destruction electronics}: self-destruction is commonly required in the military and defense applications, such as when a devices is left in the battlefield \cite{Jeremy-Hsu:2021}. In this case, PUF can inspect self-destruction signals, and only executes self-destruction if it passes authentication.
\end{itemize}

\begin{figure}[ht]
    \centering
    \includegraphics[scale=0.3]{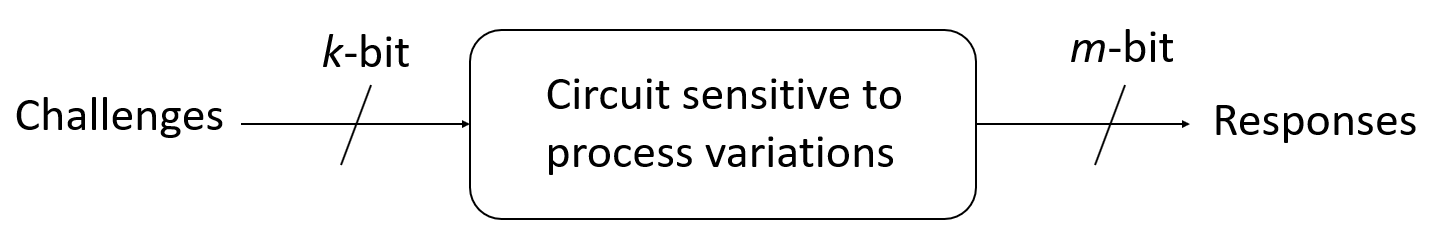}
    \caption{General operation of PUF}
    \label{fig1}
\end{figure}

PUF has been widely used to provide essential security services, such as authentication and secret key generations, specially at constrained environments, such as IoT, where power consumption and security need to be balanced. In most applications, the main function of the PUF is to authenticate IoT devices as well as store the secret keys. 
The basic operation of PUF based security is to get a random choice of challenges bits to the PUF circuit and produce unpredictable and random response. Moreover, the manufacturing process variations of the PUF circuit has a unique silicon fingerprint, which provides a unique challenge-response pair (CRP) for each IoT devices even with the same input challenge bits \cite{babaei:2019}. Figure \ref{fig1} illustrate the general operation of PUF where $k$-input bits represent the challenges and $m$-output bits provide the unique responses.

Several security criteria should be considered to achieve high secrecy. 
First, the response bits need to be correlated to the specific challenge’s bits and is reproducible for the same challenge, despite any environmental factors, such as temperature and voltage.
Second, the uniqueness of the CRP pairs should be verified by applying same challenges to different PUF circuit that must generate different responses.
Third, the challenge $k$-bits need to be large enough to limits the searchable and predictable computational of the eavesdropper.

\subsection{PUF categories}\label{PUFcategories}

Several types of PUFs have been proposed, the authors in \cite{Maes:2010} categorized PUF as: memory-based PUFs, which exploit the initial binary sequences of memory when it is powered on, and delay-based PUFs, which use delay variations between propagation signals of the circuit. However, more commonly, other authors classified PUFs into three types: weak, strong, and controlled PUFs \cite{uhrmair:2010}, each with it is own security properties and applications.

\subsubsection{Strong PUFs}
Strong PUF provides exponential growth in term of the number of CRPs pairs due to the large size of the circuits. However, while it would be impossible to efficiently recount all the CRPs pairs for large CRPs pairs, strong PUF is generally stable under environmental conditions changes. Strong PUF is also considered unpredictable, since CRPs contains multiple combination, and unclonable, where response can be readable without any additional information from the internal design of PUF. Strong PUF can be used for authentication and key establishment \cite{Schinianakis:2019}\cite{uhrmair:2010}.

\subsubsection{Weak PUFs}
Weak PUF can be utilized for key generations and digital fingerprinting. However, it accepts limited number of CRPs and is increasingly linear. Compared to strong PUF, weak PUFs provide constant response under environmental conditions, utilize small number of CRPs pairs, and provides responses that are both unclonable and unpredictable. 

\subsubsection{Controlled PUFs}\label{PUFcategories-3}
The controlled PUF use strong PUF as main block and adds a control logic to control challenges from being freely applied to the PUF circuit while preventing immediate readout of the responses. Therefore, the control logic can be utilized to hinder machine learning attack \cite{Gassend:2002}.

\section{PUF performance evaluation}\label{PUF performance}
In this section, we discuss PUF performance evaluation and quality metrics that need to be considered while designing PUF circuits to achieve high security and prevent major attacks.

A comprehensive study of PUF performance to evaluate the security can be found in \cite{Maiti:2011}. The first commercial performance evaluation based PUF-embedded for radio frequency identification (RFID) tags in \cite{Kang:2012} and the PUF performance  evaluation by delay statistics presented in \cite{Jouini:2011}.

There are four essential parameters to evaluate on a PUF circuit, namely: uniformity, uniqueness, reliability, and bit-aliasing.

\begin{itemize} 
   \item \emph{Uniformity}: the probability that the 0s and 1s are uniformly distributed in PUF's response $R$. Uniformity reflects the randomness of the response bit, and is calculated as the percentage of Hamming Weight $(HW)$ of the response bit as shown equation in (\ref{Uniformity}).
   \begin{equation} \label{Uniformity}
    Uniformity = \frac{1}{n}\sum_{l=1}^{n} r_i,_l\times100\% 
   \end{equation}
   Where \(r_i,_l\) is the \(l-th\) bit of the response $n$-bit from a chip $i$.
   \item \emph{Uniqueness}: The ability of PUF to distinguish a specific IC from other IC of the same structure when the same challenge $C$ is applied to the PUF circuit. Technically, it is defined as Inter-device (Hamming Distance) between $d$ different devices and the ideal value of uniqueness supposed to be 50\%. If the two chip $i$ and $j$ ($i \neq j$) have the responses $R_i$ and $R_j$ for the same challenge $C$, the average inter-device can be calculated as:
    \begin{equation} \label{Uniqueness}
    Uniqueness = \frac{2}{d(d+1)}\sum_{i=1}^{d-1}\sum_{j=i+1}^{d} \frac{HD(R_i R_j)}{n}\times100\% 
   \end{equation}
   \item \emph{Reliability}: The PUF design must be able to reproduce the same response bit $R$ to the same challenge $C$ under fluctuation of the environmental conditions, such as supply voltage and temperature. The reliability of PUF can be estimated as an average intra-device $(HD)$ and indicate the unreliable or noisy responses bits:
    \begin{equation} \label{HDintra}
    HD_{intra} = \frac{1}{s}\sum_{t=1}^{s} \frac{HD(R_i R_{i,t}^{'})}{n}\times100\% 
   \end{equation}
   Where $R_i$ for the chip $i$ measured at the normal operation condition and $R_{i}^{'}$ extracted at different supply voltage and temperature. $R_{i,t}^{'}$ is the $t$-th sample of $R_{i}^{'}$. The total of $n$-bit response obtained for $s$ group. In other word, the reliability is reflecting the stability of PUF and it is measured with both equations (\ref{HDintra})(\ref{Reliability}):
    \begin{equation} \label{Reliability}
    Reliability = 100\%-HD_{intra}
   \end{equation}
   \item \emph{Bit-aliasing}: Bit-aliasing indicates the similarity of PUFs responses. When bit-aliasing occurs, different IC may produce identical response. The bit-aliasing of the $l$-th bit of an $n$-bit response is the average hamming wight of the $l$-th bit across several $k$ devices. The ideal value is 50\% and it is defined as:
    \begin{equation} \label{Bit-aliasing}
    Bit-aliasing = \frac{1}{k}\sum_{i=1}^{k} r_{i,l}
   \end{equation}
   where $k$ is the number of PUF devices and $r_{i,l}$ is the $l$-th bit of the response $n$-bit response.
\end{itemize}

\section{PUF authentication protocols and key generation}\label{PUF authentication}
In this section, we describe how PUFs circuit can be used to authenticate low cost devices, such as IoT devices and Radio-frequency identification (RFID), without resorting to conventional cryptography to maintain an acceptable low power consumption and reduce the overhead circuit area. 

\begin{figure}[ht]
    \centering
    \includegraphics[scale=0.2]{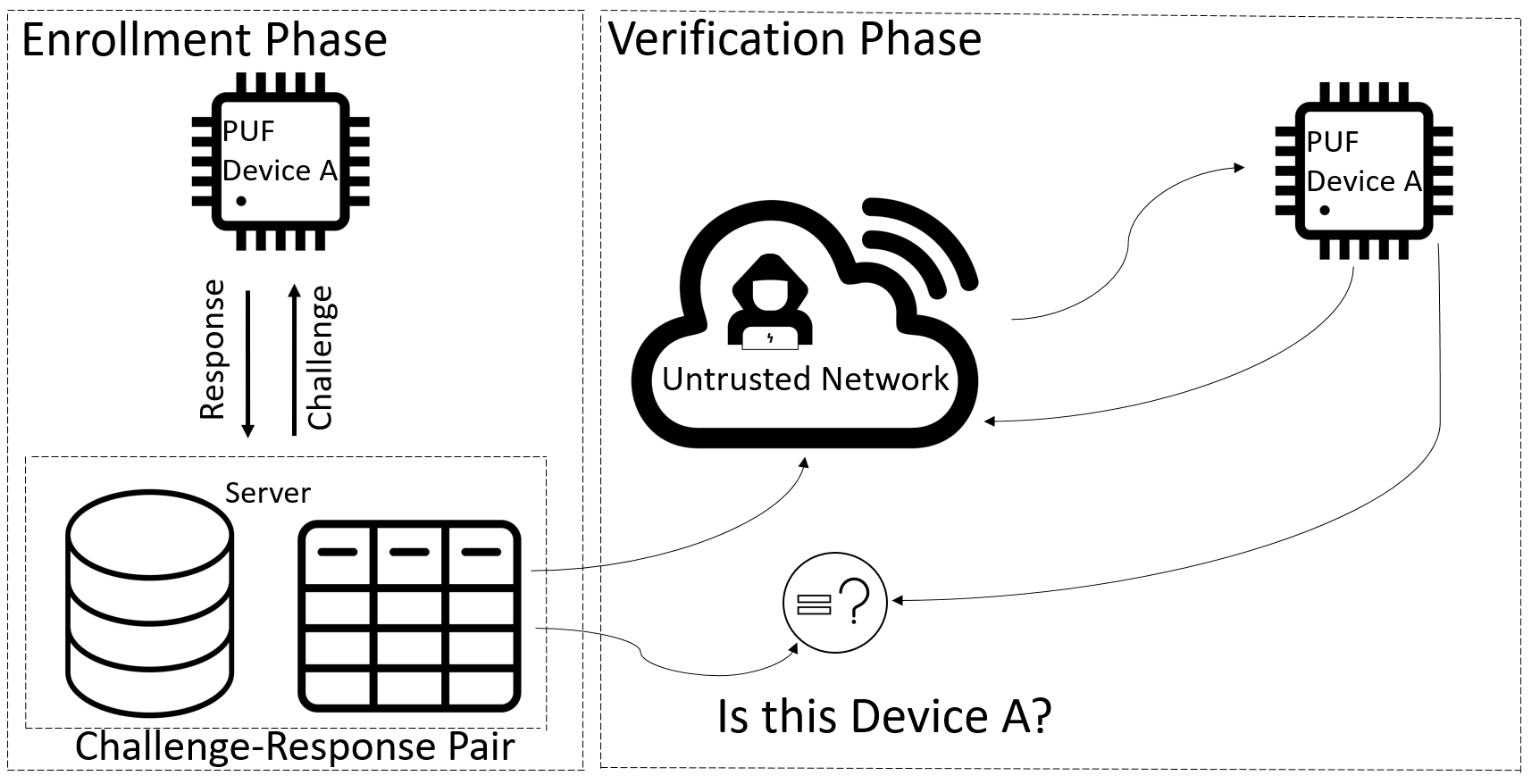}
    \caption{Authentication in PUF}
    \label{fig2}
\end{figure}

Authentication in PUF can be performed in two phases as shown in figure \ref{fig2}: the enrollment and verification phase. In the \emph{enrollment} phase, the PUF circuit is directly connected to the server to receive the challenge bits, then the PUF provides the response bits to be stored and used later in the verification phase by the server. In the \emph{verification} phase, since the PUF chip is implemented into IoT devices to be authenticated by the server, the server sends the original challenges bits that has been utilized in the enrollment phase and the IoT device reply with the generated responses bits. If the generated responses bits match any entry in the original (stored) CRPs table, the IoT devices is authenticated. Additionally, the response bits for PUF circuit can be used to extract secret key to ensure confidentiality when exchanging data \cite{babaei:2019}. Also, the challenges must never be reused to prevent man-in-the-middle attack and consequently predict the CRPs.

In fact, authentication in PUF can be done in plaintext. The authors in \cite{Suh:2007} proposed key generation protocol using PUF circuits to be implemented with conventional cryptographic primitives (e.g., RSA). The key generation operation proceeds in two steps. First, the error correction code (ECC) consisting of initialization and regeneration to ensure that PUF circuit generates the same keys under variation of environmental conditions or fluctuations of power supply and temperature. Second, key generation is executed to transform the PUF output into keys. 

Moreover, the PUF based lightweight protocol proposed in \cite{Mahalat:2018} authenticates IoT devices during an establish WiFi connection. It was shown that this protocol overcomes the security issues against several WiFi attack, such as, MAC spoofing attack, invasive attack and evil twin attack by using only 3 CRPs to secure the connections \cite{Nakhila:2016}. A mutual authentication protocol based PUF was proposed in \cite{Satamraju:2020}, which utilizes keys generated by PUF to authenticate IoT devices while using on the fly keys to avoid key storage. Furthermore, the authors in \cite{Aman:2019} introduced PUF based authentication protocol combined with exploiting wireless channel properties such as, Received Signal Strength Indicator (RSSI) to distinguish between legitimate and eavesdropper channels. Thus, data provenance is achieved in terms of confidence and data source e.g., locations and time. In addition, the authentication protocol based on continuously confirming the existence of device is proposed in \cite{Goutsos:2019}. This protocol was developed to detect the displacement of nodes through the link state changing using one CRP. Another work \cite{Noura:2019} proposed multi-factor authentication that depends on cryptographic primitive such as, hash functions and XOR gates with configurable PUF as the first factor and the second factor exploits channel characteristics such as, RSSI and Signal-to-Noise Ratio (SNR) as a fingerprint for the devices. Similar work \cite{Jiang:2019} introduced two factor authentication protocol for Internet of Vehicles (IoV). This protocol relied on a combination of password and PUF to enhance the authentication mechanism, eliminate secret key storage in devices and ensure that the adversary cannot compromise the device even with the physical access. Table \ref{tab:table2} illustrate the comparison between PUF authentication protocols.  

\begin{table*}[ht]
        \caption{PUF Authentication Protocols}
        \label{tab:table2}
     \resizebox{\textwidth}{!}{%
    \begin{tabular}{ c | m{15em} | m{10em} | m{22em} }
        \hline
        Ref & Authentication method & Technique & Comments \\  
        \hline
        \hline

        \cite{Suh:2007} & Authenticate individual ICs & PUF-based & Suitable for low-cost platform such as, RFIDs \\
        \hline
        \cite{Mahalat:2018} & Wi-Fi authentication of IoT devices & PUF-based & Less resource and computation overhead using only 3 pairs of CRPs. \\
        \hline
        \cite{Satamraju:2020} & Mutual Authentication Protocol & PUF-based & Used for real-time applications No need to store the generated keys \\
        \hline
        \cite{Aman:2019} & Authentication privacy preservation & PUF-based and wireless link fingerprint & Mitigate against physical and cloning
        attacks. Low energy consumption compared to related protocols. \\
        \hline
        \cite{Goutsos:2019} & Lightweight pairwise protocol & PUF-based & The protocol can detect nodes that have been removed or replaced \\
        \hline
        \cite{Noura:2019} & Mutual multi-factor authentication & PUF-based and cryptographic method & Lower communication overhead. Three messages required to achieve the authentication Fast to execute \\
        \hline
        \cite{Jiang:2019} & Two-factor authentication for IoV system & PUF-based and cryptographic method & There is no storage required of
        any secret data. They combine password with PUF (two-factor) \\
        \hline
       
    \end{tabular}%
    }

\end{table*}

\section{PUF Architectures}\label{PUF Architectures}
In this section, we describe several PUF architectures suitable for IoT applications. We will also discuss the strengths, weakness, quality metrics and evaluation of common architectures as shown in table \ref{tab:table3}. The following criteria need to be taken into account when selecting a PUF architecture:  

\begin{enumerate}
   \item Robustness against different possible attack, such as machine learning attack and side channel attack \cite{babaei:2019}.
    \item Statistical properties and quality metrics such as:
        \begin{itemize}
            \item \emph{Uniqueness}: the ability of PUF circuit to generate a unique secret key when a challenges bit is provided.
            \item \emph{Reliability}: the ability of the PUF circuit to generate the same secret key under different environmental factors, e.g., temperature and voltages.
            \item \emph{Randomness}: the response bits generated from PUF circuit contain sufficient entropy.
        \end{itemize}
   \item The growth of the number of CRPs in strong and weak PUF need to be taken into account as it can lead to increased computational complexity, which, in turn, will consume power.
   \item The PUF circuits need to be implemented easily in silicon chip.
\end{enumerate}

\subsection{Arbiter PUF}\label{ArbiterPUF}
\begin{figure}[ht]
    \centering
    \includegraphics[scale=0.3]{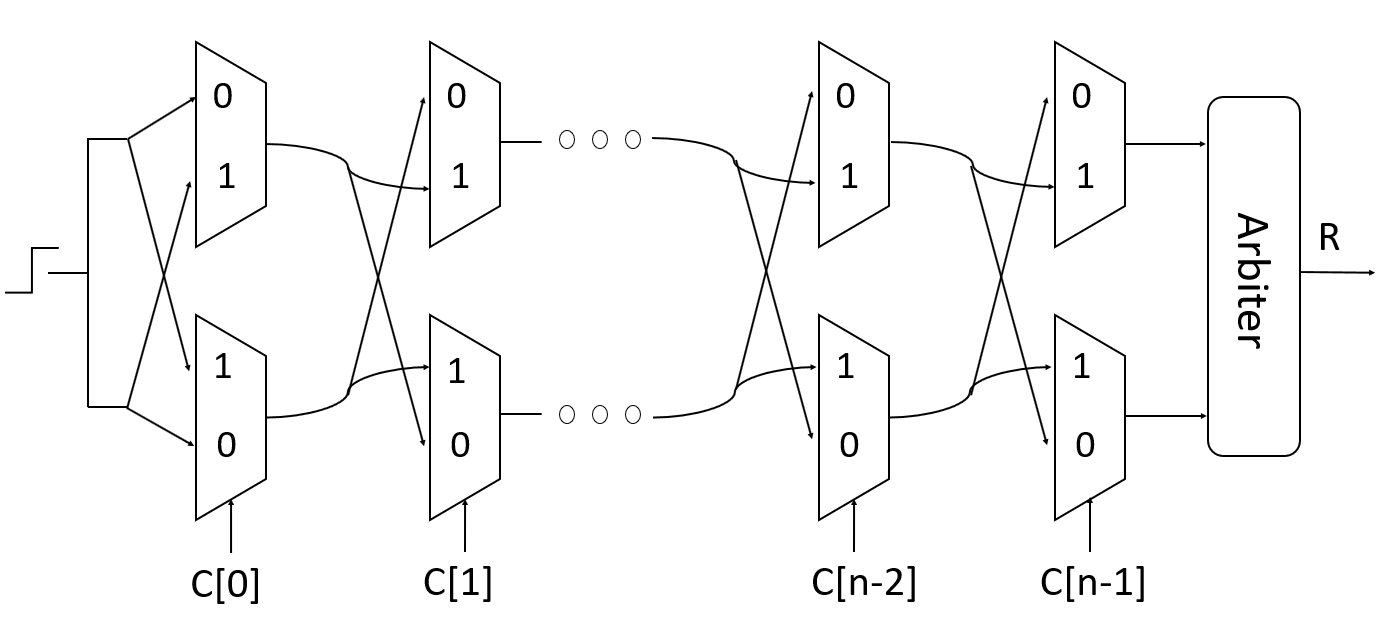}
    \caption{Arbiter PUF}
    \label{fig3}
\end{figure}

During manufacturing variations of multiplexers, different delay paths are formed, where one path is usually faster than others. Depending on the input challenges bits, each multiplexer will select the next path to be switched to, which provides multiple combinations of bit path selections. Arbiter PUF \cite{Gassend-B:2002} operates by comparing two path delays as shown in figure \ref{fig3} and generates a response bit ‘0’ or ‘1’ depending on the faster path being selected by the latch at the output.

Arbiter PUF is categorized as strong PUF. Moreover, to achieve practical statistic properties, all the delay-paths must have the same length. The arbiter PUF can be implemented in both Field Programmable Gate Arrays (FPGA) and application-specific integrated circuit (ASIC).   

In \cite{Dubrova:2018}, a reconfigurable arbiter PUF presented with 4$\times$4 switch block instead of the classical 2$\times$2. such that the 4$\times$4 switch block can be reconfigured to increase the numbers of paths connection, which can be used for the applications that required regular keys generation. The authors in \cite{Cao:2019} proposed an energy efficient arbiter PUF using Current starved (CS) inverters at the back stage of each multiplexers. The proposed arbiter utilizes two RS latches and NAND gate instead of D flip-flop in classical arbiter to improve the propagation delay at the output phase. Therefore, the design alleviates effectiveness of fluctuating temperature at the output responses. Similarly, in \cite{Moradi:2020}, the authors demonstrated an energy efficient arbiter PUF that consists of 64 PUF cells using 45nm CMOS technology, each cell contains 8 switching elements which competes between different paths depend on challenge $C$ values, 8 selecting modules, and an arbiter. The design achieves high uniqueness and consume low energy.

\subsection{Ring Oscillator PUF}\label{ROPUF}
\begin{figure}[ht]
    \centering
    \includegraphics[scale=0.3]{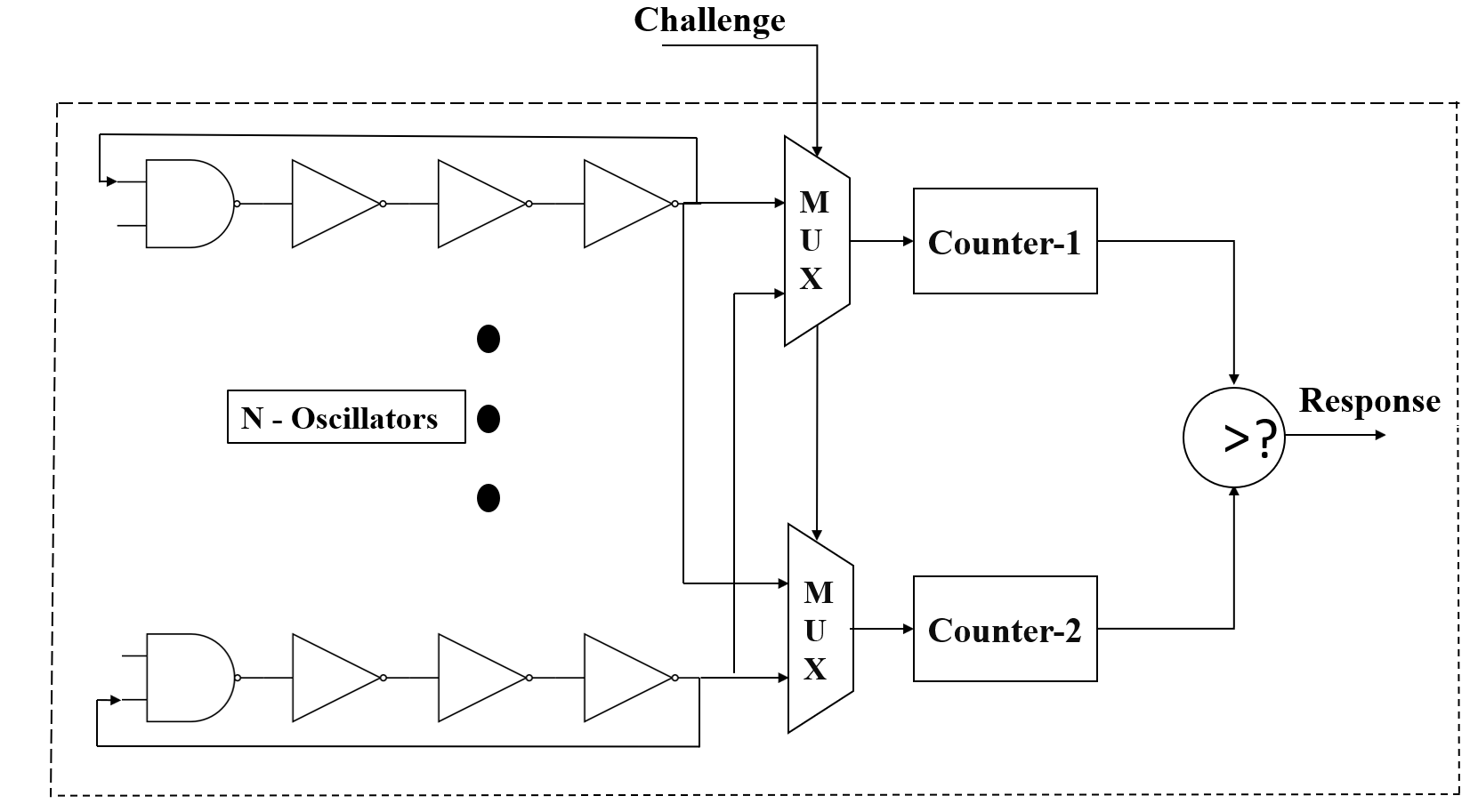}
    \caption{Ring Oscillator PUF}
    \label{fig4}
\end{figure}

Ring Oscillator (RO) PUF \cite{Suh:2007} is based on the circuit oscillation between two voltage levels in specific frequency as shown in figure \ref{fig4}. By comparing two RO frequencies, the binary bits are generated based on incoming challenges bits. However, while the theoretical properties of RO PUF show that the oscillating frequencies must be the same, during the hardware manufacturing variations process it will inevitably cause some differences in the oscillation frequencies. 

RO PUF is a strong PUF and can be implemented on FPGA. The main drawback of RO PUF is its sensitivity to the environment. To address this issue, the authors in \cite{Deng:2020} proposed configurable RO using only two hybrid logic gates, which is not only reliable under environment variation conditions, but also consumes less power and circuit area. The work in \cite{Qian:2019} further enhances the response entropy by adding configurable multiplexer with RO PUF circuit, which can select from inputs challenges $C$ based on proposed selection algorithm. The FinFET 20 nm technology based RO PUF proposed in \cite{Zayed:2019} to overcome hardware overhead and power consumption of classical RO PUF, however they introduce frequency divider with flip-flop instead of counters, comparator to reduce power consumption. 

\subsection{SRAM PUF}
\begin{figure}[ht]
    \centering
    \includegraphics[scale=0.5]{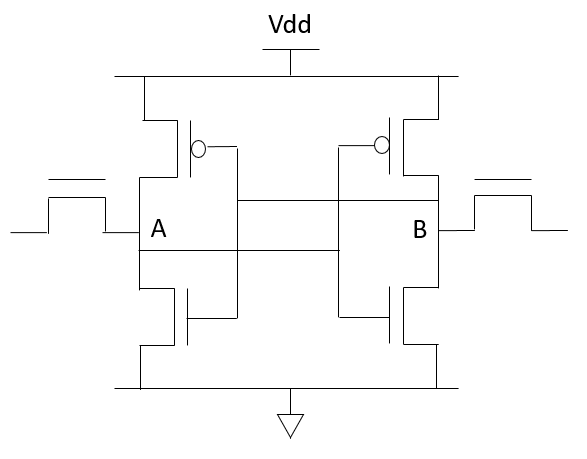}
    \caption{SRAM PUF}
    \label{fig5}
\end{figure}

One of the common PUF based on memory architecture is SRAM PUF \cite{D.E.Holcomb:2007} \cite{Guajardo:2007}. The main idea of SRAM PUF is to generate a response bit based on boot-up of SRAM cells, which are unpredictable; that is, when the SRAM is powered ON, the initial values of the single cells in the SRAM can be ‘0’ or ‘1’ randomly as they are considered noisy fingerprint. Furthermore, each SRAM has unique states during this boot-up period. Figure \ref{fig5} illustrates SRAM PUF circuit. 

However, SRAM PUF is weak PUF with limited number of CRPs and is mainly applicable for microcontrollers. Hence, more layers of security are required to thwart machine learning attacks. Additionally, SRAM PUF suffers from noise effects and requires error correction \cite{babaei:2019}\cite{D.E.Holcomb:2007}.

\subsection{Hybrid PUF}
\begin{figure}[ht]
    \centering
    \includegraphics[scale=0.3]{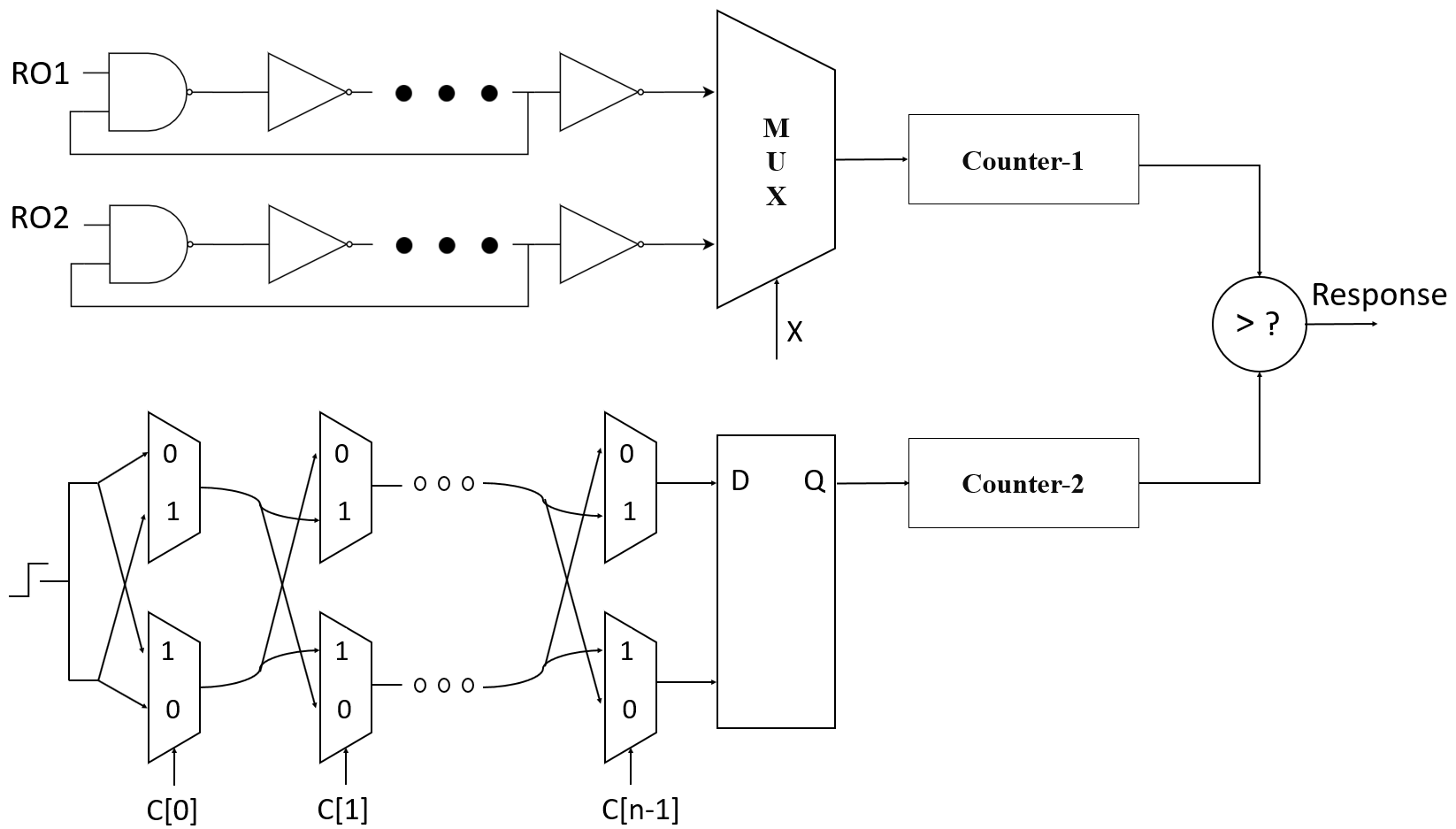}
    \caption{LH-PUF}
    \label{fig6}
\end{figure}

Lightweight Hybrid PUF (LHPUF) \cite{Sankaran:2018} combines features of Arbiter PUF and RO PUFs to enhance the security as illusrated in figure \ref{fig6}. LHPUF consists of $N$ to 1 multiplexers, two counters, NAND gate, NOT gate, and one arbiter circuit. The result of bit output response depends on the count of the number of ‘1’s or ‘0’s in the counter at the output. The authors in \cite{Sankaran:2018} implemented LH-PUF using FPGA (Xilinx) and it provides higher security performance as shown in table \ref{tab:table3} compared to traditional arbiter PUF and RO PUF with less power consumption.

Another work \cite{Yanambaka:2016} introduced FinFET based-PUF with two hybrid oscillator-arbiter PUF designs, which improved the power consumption and speed compared to traditional RO and arbiter PUF.

\subsection{Optical PUF}
The optical PUF was first proposed in \cite{Pappu:2001}\cite{Pappu:2002} to demonstrate an inexpensive non-silicon system that consists of a token with the integrated three-dimensional micron scale glass as a physical system to generate 2400-bit unique key. The authors in \cite{michael:2017} developed an optical PUF that can be implemented in Printed Circuit Board (PCB) by adding an imager and Light-emitting diodes (LED). These components are covered by polymer waveguide. However, the LED light is emitted and reflected by the waveguide to the imager to generate a unique number that can be used for authentication and key generation. Therefore, any invasive attack attempting to discover the unique key will destroy the waveguide coating, which  damages the secret key. Rührmair et al \cite{Rhrmair:2013} represent new image transformation that enhance the PUF entropy to measure the interference pattern through the optical PUF instead of detecting the reflection compared to the previous work in \cite{Pappu:2001}\cite{Pappu:2002} with the same hardware cost. 

\subsection{Memristor PUF}
Memristor (refer to memory resistor) is first proposed in 1971 by Chua \cite{Chua:1971}, which provides a relation between charge and flux. In 2008, HP labs \cite{strukov:2008} present a physical model of two-terminal device, which is based on switching process between two resistance states, such as high resistance state (HRS) (called OFF state, ‘0’) and low resistance state (LRS) (called ON state, ‘1’). This state is changed when the voltage is applied to across terminals for specific period with two main operations: SET (that represent the change of state from HRS to LRS) and RESET (refer to transition from LRS to HRS). As a result, memristor is suitable for PUF due to the variability of the state and switching process. Also, it is used as True Random Generator \cite{Uddin:2019}. Koeberl et al \cite{Koeberl:2013} proposed memristor PUF to exploits undefined logic state region based on memory functionality that depends on access time and applied voltage. Consequently, unpredictable sequence value is produced due to the utilization of the weak-write method. The authors in \cite{Rose:2013} present a single bit memristor PUF with two control signals that specify the writing and reading operation. On the other hand, the authors in \cite{RoseA:2013} describe a multi bit memristor PUF as an entropy source based on the process variations of the write-time memristor cell. The Xbar memristive architecture \cite{Uddin:2018} that consists of $N\times M$ size word-lines as a rows to receive the challenge bits $N$ and bit-line as a columns to produce response bits $M$. The more Xbar rows size, the more randomness generated. An optimized and robust architecture based on memristive Xbar is presented in \cite{Muhammad:2021}\cite{Khan:2021}, which implements two memristive Xbar PUF utilizing two memristors devices and the logic circuit that processing the challenges and activate the PUFs. 

\subsection{Quantum PUF}
Quantum PUF provides control on the unique parameters of the process variation that created the classical PUF. {\v{S}}kori{\'{c}} in \cite{Boris:2010} proposed quantum readout PUF (QR-PUF), which uses quantum state for both challenges and responses, usually to implement remote authentication protocols. By exploit the no-cloning theorem \cite{Wootters:1982} of the quantum state, the adversary will be detected if they intercept the CRP. Another work \cite{Goorden:2014} present quantum secure authentication (QSA), which rely on phase shaping of irradiate light pulse using spatial light modulator (SLM) and analyzer plane to detect the reflected response. The quantum confinement is described in \cite{roberts:2015} to provide a unique identifier for the devices by measuring the variation in resonant tunnelling diodes (RTD). A comprehensive study of the quantum PUF is presented in \cite{arapinis:2021}, which defines a quantum attack model and security parameters of the quantum PUF. The authors in \cite{Phalak:2021} propose a quantum PUF to address the security issues of a workload scheduling algorithm threats for cloud-based quantum computers.   

\subsection{MRAM PUF}
Magnetoresistive Random-Access Memory (MRAM) based PUF is proposed in \cite{Jayita:2014} \cite{Das:2015} to generate unique keys and provide authentication by exploit the variation geometric of the MRAM cell. The stream bits are stored in Magnetic Tunnel Junction (MTJ) that consists of several layers. Due to the variation of the manufacturing process, the geometric of the cell vary in shape (rectangle or ellipse). Experimental evaluation of the MRAM PUF is presented in \cite{Nejat:2020} with Thermally Assisted Switching MRAM (TAS-MRAM) method that are fabricated in dies. The experiment shows that TAS-MRAM consumes low power and high speed compared to SRAM. Furthermore, the authors in \cite{Ali:2021} proposed reconfigurable arbiter PUF based on hybrid Spin Transfer Torque (STT-MRAM/CMOS). The design employed the variation process of the transistors and MTJs connected in series with the control signals. The switch selection and pre-charge sense amplifier (PCSA) is used as an arbiter to determine the delays of the discharge current between paths. Therefore, the design provide sufficient entropy response and produce large and unpredictable CRPs compared to the silicon/classical arbiter PUF (section \ref{ArbiterPUF}). The reliability enhancement/improvement of STT-MRAM PUF response is demonstrated in \cite{Hu:2021}.

\subsection{Carbon PUF}
Carbon Nanotube Field Effect Transistor (CNTFET) based PUF is a promising technology to provide a unique signature with low power consumption. The first design of Carbon Nanotube PUF (CNPUF) is presented in \cite{Konigsmark:2014}, which is composed of pairs of CNTFET connected in series that share the same input voltage and response bits produced from comparing two stage currents. The simulation result in \cite{Konigsmark:2014} shows that the CNPUF is reliable under temperature and voltage variation. The authors in \cite{Moradi:2017} introduced CNTFET PUF cell, which compares the input voltage twice using two inverters and comparator. Similarily, the work in \cite{Lee:2019} demonstrated the fabrication of 400 CNTFET PUF device with same manufacturing process and evaluate their performance. The measurement show that the devices produce high quality metrics in term of uniformity and uniqueness. Ternary cycle operator based CNTFET PUF is proposed in \cite{Srinivasu:2021} with two delays line using cycle operators. 

\subsection{Others PUFs}
Several other PUF designs have been proposed to enhance the quality metrics, including: 
\begin{itemize} 
   \item \emph{Glitch PUF}\cite{Suzuki:2010}: exploits the glitch waveform variation of the logic gates.
   \item \emph{Butterfly PUF}\cite{Kumar:2008}: overcomes the initialization SRAM in some FPGAs making it suitable to be implemented on all FPGAs types.
   \item \emph{Latch PUF}\cite{Su:2007}: introduce a unique identification number (ID) for IC using cross-coupled NOR gates arrays, which improve the speed and power consumption.
   \item \emph{Digital PUF (D-PUF)}\cite{Miao:2018}: improves the reliability of analog PUFs.
   \item \emph{Coin Flipping PUF (CF-PUF)}\cite{Tanaka:2018}: exploits the convergence time of the bistable ring circuit.
   \item \emph{Finite State Machine (PUF-FSM)}\cite{Gao:2018}: removes the need of error correction code (ECC) in Controlled PUF (section \ref{PUFcategories-3}) and improve the security.
   \item \emph{Subthreshold current array PUF (SCA-PUF)}\cite{Zhuang:2020}: exploits the $I-V$ characteristics of the two arrays of transistors and the response is produced based on the comparison between two output voltages.
   \item \emph{Spin Orbit Torque (SOT) PUF}\cite{Cao:2021}: reconfigurable PUF based on Spin Orbit Torque (SOT) to stimulate the motion of the Domain Wall (DM). 
\end{itemize}

\section{PUF implementation}\label{PUF implementation}
FPGA is widely used to simulate the design of PUF circuit due to flexibility, customizability and configurable logic gate as well as faster to be deployed in IoT devices. In this section, we discuss PUF implementation in FPGAs. We will cover only the architectures that are suitable for FPGA implementation. Table \ref{tab:table3} summarizes the quality metrics and evaluation of designs in this section.

\subsection{Arbiter PUF on FPGA}
Many types of Arbiter FPGA PUFs has been proposed to improve the security of lightweight devices. The authors in \cite{Machida:2014} proposed 3-1 Double Arbiter PUF with 3 arbiter PUF and 1 bit response and new mode of operation for wires connection to arbiter which enhance uniqueness of output responses. The feedback Oriented XOR flip-flop based arbiter (FOXFF APUF) \cite{Sushma:2018} for identification applications and provides a uniqueness improvement compared to FFAPUF \cite{Gu:2016} by adding delay elements, such as, feedback flip-flop to the design. Another work \cite{Sahithi:2018} introduced a combination of flip-flop and XOR gates based arbiter which enhance the uniqueness to 16\% using families of FPGAs and consume more resources compared to \cite{Gu:2016} design. Flip-Flop Based Arbiter PUF (FF-APUF) demonstrated in \cite{Gu:2019} provides sufficient entropy and reliability compared to conventional arbiter PUF, which is suited for FPGA implementation and can be utilized in authentication protocols for lightweight devices. 
Moreover, the authors in \cite{He:2020} introduced the new concept of bit self test (BST) based arbiter PUF by designing a detection circuit that produces reliability flag. BST depends on propagation delay between paths and fluctuations in temperature and voltages. Therefore, based on the reliability flag a robust responses generated, which is suitable for key generation and authentication.

\subsection{RO PUF on FPGA}

One of the alternative solution to conventional RO PUF is Transient effect ring oscillator (TERO-PUF) \cite{Bossuet:2014}, which exploits oscillation of four 64-loop TERO cells; each cell consists of cross-coupled circuit of 2 AND gates and 2 inverters. The design thwarts electromagnetic attack \cite{Bayon:2013}, which analyses electromagnetic emanation and obtains information from RO circuit. However, the circuit utilizes more hardware resources. Furthermore, the authors in \cite{Yan:2017} introduce Phase Calibration Process (PCPUF) technique to precisely measure the frequency of 128 RO array, which improves the stability and reduces the bit error rate of responses. In addition, the authors in \cite{Garcia-Bosque:2020} demonstrated Galois ring oscillators (GARO-PUF) that compares different statistical parameters, such as, variability and location of implemented PUF in FPGA of oscillators instead of frequencies, so the design overcomes the systematic issues that produced by RO frequencies and correlations when RO PUF implemented in some physical locations in FPGA. The RO PUF based Lookup Table (LUT) FPGA introduced in \cite{Li:2020} extracts more entropy by applying proposed method called Difference on Summed Difference (DSD) to obtain the differences between frequencies. The design achieved sufficient entropy with low area overhead. Moreover, the authors in \cite{Aguirre:2020} proposed two schemes: parallel and serial based RO PUF, and replace the counters with Linear feedback shift register (LFSR) to eliminate the linear behaviour in counters. However, the problem of linearity still appear in LFSR, so the scrambler circuit is also proposed to obliterate the correlation behaviour. Therefore, the design produces unpredictable output and consumes low area and power. In \cite{Yao:2021}, a reconfigurable XOR gate based RO PUF is proposed, which produced larger CRP and enhance responses stability; reconfigurable XOR gate implemented in the RO circuit instead of inverter.

\subsection{LFSR PUF on FPGA}

Generally, Linear feedback shift register (LFSR) has been widely used as Random Number Generator (RNG) in cryptography, specially for lightweight devices with limited hardware resources. One of security concern associated with LFSR is its linearity and predictability. To address this, the authors in \cite{hou:2020} proposed lightweight configurable Shift Register based PUF (SRPUF), with a non-linear function to improve the entropy and thwart machine learning attacks. Similarly, the work in \cite{Hou:2019} introduced LFSR-based strong PUF (L-PUF), which is a weak PUF at front-end of the circuit, such as Anderson PUF \cite{Anderson:2010}, combined with LFSR. Consequently, the authors in \cite{Amsaad:2018} proposed asynchronous LFSR based PUF (LFSR-PUF) that utilizes basic building block of the FPGA such as, LUTs and flip-flops, which exploits the variation process of LFSR circuit to produce random responses. Another work \cite{Zhou:2020} proposed pseudo linear feedback shift register with multiple RO (PL-MRO) PUF, which utilizes logic gates instead of shift registers in LFSR to exploit the delay behaviour of the RO circuit. The PL-MRO PUF, produces sufficient entropy, high speed operation and low power consumption compared to conventional RO PUF.

\subsection{Others PUF on FPGA}

Several other PUF architectures have been proposed targeting FPGAs implementation and taking advantages of inherent FPGAs structures such as, LUT and flip-flops. For instance, the authors in \cite{Ardakani:2016} proposed an area efficient SR-Latch PUF with two methods of implementations that consist of 4 NAND gates with multiplexers to generate high entropy responses. In addition, the work in \cite{Wang:2018} introduced parallel scan design based PUF and exploits the delay difference between pairs of shift registers as chains through the SR-Latch arbiter to reduce the area overhead and improve the uniqueness. Consequently, a combination of weak PUF and pseudo Strong PUF (p-SPUF) were proposed in \cite{Shen:2019}. The weak PUF produces 1-bit response from the variation process of the logic gates, which feeds as an input to the LFSR. The design enhances the response randomness and can be well fitted in FPGAs due to low area cost. Furthermore, the dynamic reconfigurable PUF introduced in \cite{Cui:2019} based on 3 different logic circuit design stored in external memory, so the FPGA can be configured by the programming system (PS) that has an access to external memory to select between configurable logic. This technique improves the hardware overhead, provides large amount of CRPs and thwarts machine learning attack. Similar work \cite{Wei:2020} proposed transformer PUF based on reconfigurable properties of RO basic circuit such as, multiplexers which select between different paths and configurable XOR gates. As a result, transformer PUF improves hardware efficiency and reliability compared to conventional configurable RO. Moreover, a modified Anderson PUF with Low-density parity checker error correction (LDPC) is proposed in \cite{kalya:2020} to enhance the error bits in responses. The LDPC was utilized to provide reliability under environmental variations condition and high uniformity. However, the design has less uniqueness which is not suitable for authentication applications. Another work \cite{Lotfy:2021} introduced optimization of Anderson PUF that utilized one configurable logic block (CLB) in FPGA with inherent XOR gates, which improves unpredictability of responses compared to conventional Anderson PUF.

\begin{table*}
    \begin{adjustbox}{max width=\textwidth, max totalheight=\textheight}
    \begin{threeparttable}
    \caption{Summery of PUF quality metrics and evaluation.}
    \label{tab:table3}
    \begin{tabular}{ c | c | c | c | c | c | c | c } 

        \hline
        \makecell{Ref} & \makecell{PUF \\ Architectures} & \makecell {Implementation \\ Platform} & \makecell{Uniqueness \\ (inter-chip) \\ $HD$} & \makecell{Uniformity \\$HW$} &  \makecell{Reliability \\ (intra-chip) \\ $HD$} & \makecell{Hardware \\ Overhead} & \makecell{Power \\ Consumption} \\ 
        \hline
        \hline
        \cite{Lee:2004} & Arbiter PUF & TSMC 0.18 $\mu$m & 23\% & --- & 95.18\%\tnote{a}, 96.26\%\tnote{b} & 1212$\mu$m$\times$1212$\mu$m & 137$\mu$W \\ 
        \cite{Daihyun:2005} & Feed-Forward Arbiter & TSMC 0.18 $\mu$m & 38\% & --- & 90.16\% & --- & ---  \\ 
        \cite{Machida:2014} & 3-1 DAPUF & Virtex-5 & $\approx$50\% & $\approx$50\% & $\approx$88\% & --- & --- \\
        \cite{Sushma:2018} & FOXFF APUF & Spartan-3, Virtex-6 & 42\%, 44\% & --- & --- & 703 Slices & --- \\
        \cite{Sahithi:2018} & XOR FFAPUF & 3 Types FPGAs & 48\%\tnote{g} & --- & --- & 224 Slices\tnote{g} & --- \\
        \cite{Dubrova:2018} & Reconfigurable APUF  & Simulation & 50\% & 50\% & --- & 4$\times$4 switch blocks & --- \\
        \cite{Cao:2019} & CS-APUF & CMOS & 46.8\% & Pass NIST & 99.2\% & 3838 $\mu$m\(^2\) & 68.63$\mu$W \\
        \cite{Gu:2019} & FF-APUF & Artix-7 & 41.53\% & 54\% & 97.10\%\tnote{a}, 93.90\%\tnote{b} & 128 Slices & --- \\
        \cite{He:2020} & BST-APUF & Artix-7 & 49.1\% & 50.3\% & $\approx$100\%  & 150 Slices & --- \\
        \cite{Moradi:2020} & Energy-Efficient APUF & CMOS & 49.9\% & 50\% & 94.4\%\tnote{a}, 97\%\tnote{b} & 2168$\mu$m\(^2\) & 6.63mW \\
        \cite{Suh:2007} & RO PUF  & Virtex-4 & 46.15\% & --- & 99.52\% & 16 $\times$ 64 array\tnote{c} & ---  \\
        \cite{Bossuet:2014} & TERO-PUF & ALTERA DE1 & 48\% & --- & 98.3\% & 416\tnote{r} & --- \\
        \cite{Deng:2020} & Configurable RO &  Artix-7 & 56.1\% & 50.36\% & 98.22\tnote{a}, 99.56\tnote{b} & 1128 $\mu$m\(^2\) & 119.4$\mu$W \\
        \cite{Li:2020} & RO based LUT & Kintex-7 & 49.15\% & --- & 99.16\% & 8\tnote{s} & --- \\
        \cite{Aguirre:2020} & Two schemes RO PUF & Spartan-7 & 46.4\% & High entropy & 99.68\% & 32 RO array (serial) & --- \\
        \cite{Yao:2021} & XOR-RO & Virtex-6 & 48.4\% & Pass NIST & 98.33\%\tnote{a}, 98.25\%\tnote{b} & 16\tnote{s} & --- \\
        \cite{Hou:2019} & L-PUF & Alinx & 49.66\% & 49.8\% & --- & 210\tnote{s} & --- \\
        \cite{hou:2020} & SRPUF & Alinx & 49.9\% & 49.8\% & $\approx$100\% & 20,544 \tnote{s} & 0.242W \\
        \cite{Amsaad:2018} & LFSR-PUF & Spartan-3E & 47.4\% & --- & 95.8\%\tnote{a}, 93.55\%\tnote{b} & 2 Slices & --- \\
        \cite{Zhou:2020} & PL-MRO PUF & Artix-7 & 51.7\% & High entropy & 94.5\% & --- & 23.44$\mu$W \\
        \cite{Anderson:2010} & Anderson PUF & Virtex-5 & 47.9\% & --- & 96.4\% & 2 Slices & --- \\
        \cite{Ardakani:2016} & SR-Latch PUF & Spartan-3 & 49.2\% & 54.27\% & 80\% & 2 Slices & --- \\ 
        \cite{Wang:2018} & Scan chains PUF & Virtex-5 & 49.86\% & Pass NIST & 96\%\tnote{a}, 98\%\tnote{b} & 128 Slices & --- \\
        \cite{Shen:2019} & p-SPUF & Alinx & 50.58\% & 49.6\% & 93.3\%\tnote{a} \tnote{b} & 12$\times$32 slices & --- \\
        \cite{Wei:2020} & Transformer PUF & Artix-7 & 49.44\% & --- & 98.12\% & Efficient & --- \\
        \cite{Lotfy:2021} & Modified Anderson PUF & Spartan-6 & 50.89\% & 49.41\% & 91.25\%\tnote{a} & 348 Slices & --- \\ 
        \cite{Yan:2017} & PC-PUF & Kintex-7 & 50\% & Pass NIST & $\approx$99\% & 365 Slices & --- \\
        \cite{Zayed:2019} & FDRO-PUF & Simulation & 49.98\% & 50.05\% & 99.5\%\tnote{a}  \tnote{b} & --- & 264$\mu$W \\
        \cite{Garcia-Bosque:2020} & GARO-PUF & Pynq Z2 & 39.1\% & $\approx$50\% & 98.9\% & 7 LUT\tnote{s} & --- \\
        \cite{Guajardo:2007} & SRAM PUF  & FPGA & 49.97\% & --- & 88\%\tnote{a} & 4600 SRAM memory bits & ---  \\
        \cite{Sankaran:2018} & LH-PUF  & Spartan-3E, Spartan-6 & 38.28\% & --- & --- & 123 slices & 2500$\mu$W \\
        \cite{Yanambaka:2016} & Hybrid PUF & Simulation & 50.9\%\tnote{p}, 52\%\tnote{q} & --- & 99.21\% & --- & 285.5$\mu$W\tnote{p}, 320$\mu$W\tnote{q} \\ 
        \cite{Pappu:2002} & Optical PUF  & Optical token &  49.79\% & --- &  25.25\% & 1cm$\times$1cm$\times$2.5mm\tnote{d} & --- \\
        \cite{Rhrmair:2013} & Optical PUF  & Optical token &  50\% & --- &  94\% & 1cm$\times$1cm$\times$2.5 mm\tnote{d} & --- \\
        \cite{Su:2007} & Latch PUF  & CMOS 0.130$\mu$m & 50.55\% & --- & 96.96\%\tnote{b}, 96.22\%\tnote{b} & 8$\times$16 NOR-latch array & 0.162$\mu$W \\
        \cite{Koeberl:2013} & Memristor PUF  & Fabricated & High & $\approx$50\% & --- & 150(1MB)\tnote{e} & --- \\
        \cite{RoseA:2013} & Memristor PUF  & Simulation & High & $\approx$50\% & --- & --- & --- \\
        \cite{Uddin:2018} & Memristor PUF  & Simulation & $\approx$50\% & $\approx$50\% & 90\%\tnote{a} & Four\tnote{f} & 104$\mu$W \\
        \cite{Muhammad:2021}\cite{Khan:2021} & Memristor PUF  & Simulation & 50.1\% & 48.9\% & 97\%, 95.5\% & Two 32$\times$2\tnote{h} & --- \\
        \cite{Phalak:2021} & Quantum PUF  & Cloud-based\tnote{i} & 55\% & --- & 96\% & --- & --- \\
        \cite{Jayita:2014} & MRAM PUF  & Fabricated & 47\% & --- & 99.9\% & 10$\times$20 PUF array\tnote{j} & --- \\
        \cite{Nejat:2020} & MRAM PUF  & Fabricated\cite{crocus:2021} & 49.8\% & --- & 92.3\%\tnote{b} & 32$\times$32 PUF array & --- \\
        \cite{Ali:2021} & MRAM PUF  & CMOS & $\approx$50\% & Pass NIST\tnote{k} & 99.76\%\tnote{a} \tnote{b} & 28nm & --- \\
        \cite{Hu:2021} & MRAM PUF  & CMOS & 50.64\% & 50.02\% & 97.87\%\tnote{a}   \tnote{b} & $m\times n$ PUF array\tnote{l} & --- \\
        \cite{Konigsmark:2014} & Carbon PUF  & Simulation & 49.67\% & --- & 96.5\%\tnote{a} \tnote{b} & serial connection\tnote{m} & 1.26$\mu$W  \\ 
        \cite{Moradi:2017} & Carbon PUF  & Simulation & $\approx$49\% & $\approx$50\% & 96.67\%\tnote{a}, 80.41\%\tnote{b} & 1.27$\mu$m\(^2\)\tnote{n} & 1350$\mu$W  \\
        \cite{Lee:2019} & Carbon PUF  & Fabricated & 49.7\% & 45.3\% & --- & 500$\mu$m PUF array & --- \\
        \cite{Srinivasu:2021} & Carbon PUF  & Simulation & --- & 48.7\% & --- & --- & 7.26$\mu$W \\
        \cite{Suzuki:2010} & Glitch PUF  & Spartan-3A & 41.5\% & --- & 93.4\% & --- & --- \\
        \cite{Kumar:2008} & Butterfly  PUF  & Virtex-5 & $\approx$50\% & --- & 94\% & 130 slices & --- \\
        \cite{Miao:2018} & Digital PUF  & Simulation & $\approx$50\% & 50\% & $\approx$100\%\tnote{a} \tnote{b} & 145.92$\times$89.32 $\mu$m\tnote{o} & --- \\
        \cite{Zhuang:2020} & SCA-PUF  & CMOS & $\approx$50\% & 52.8\% & 99.9\%\tnote{a} \tnote{b} & 130nm & 68nW \\
        \cite{Cao:2021} & SOT PUF  & Fabricated & 47.2\% & 48.6\% & stable & 6$\times$6$\mu$m\(^2\)(device size) & --- \\
        \hline

    \end{tabular}

    \begin{tablenotes}
        \item[a] Temperature variation.
        \item[b] Voltage variation.
        \item[c] 16 $\times$ 64 array = 1024 ROs; One RO (contain 5 inverters and 1 AND gate).
        \item[d] Plastic token size.
        \item[e] Memristor memory structure.
        \item[f] Different crossbar sizes (XORed Xbar) 4$\times$2, 8$\times$2, 16$\times$2, 32$\times$2.
        \item[g] Spartan-6.
        \item[h] Memristive Xbar PUFs.
        \item[i] IBM quantum hardware.
        \item[j] The array fabricated using standard electron beam lithography process.
        \item[k] National Institute of Standards and Technology, test results for 10K responses produced from PUF.
        \item[l] MRAM Size (m$\times$n cells), 4$\times$128, 8$\times$128, 16$\times$128, 32$\times$128.
        \item[m] Serial connection of Carbon Nanotube PUF Parallel-elements.
        \item[n] The layout sketched by CAD tool Electric.
        \item[o] Overall silicon area (chip dimension). 
        \item[p] Power Optimized Hybrid Oscillator-arbiter design.
        \item[q] Speed Optimized Hybrid Oscillator-arbiter desgin.
        \item[r] Logic Array Block. 
        \item[s] Lookup Table.
    \end{tablenotes}

\end{threeparttable}
\end{adjustbox}

\end{table*}

\section{Threat landscape and Security}\label{Threat landscape}

In this section we discuss some attacks and threats on IoT devices based on PUF. We start by describing some possible attacks and assumption. Then we consider the following scenarios: 
\begin{itemize}
    \item Scenario 1: an adversary eavesdropping the communications channel between IoT devices.
    \item Scenario 2: an eavesdropper with physical access the devices.
\end{itemize}

In addition, the adversary can be active or passive; an active adversary can manipulate the operational temperature and power supply, while a passive adversary attempts to observe and intercept data in the communication channel. Ultimately, the physical access to the PUF chip is required for active attacks.  

\subsection{Brute-force} 

The attacker can try to clone the CRP table of the PUF circuit by enumerating all possible combination of challenge-response pairs by repeatedly querying the PUF circuit. Clearly, this attack is very time consuming and will generate a huge CRP table requiring a considerable storage.

\subsection{Invasive attack }

This attack measures internal PUF properties, such as delays of the circuit, to predict the response for a particular challenge. Such measurement entails physically removing external packaging and chip metal layers. However, as any probing attempt will directly affect the wiring and routing of the circuit's delayed-paths, such actions will inevitably change the PUF chip characteristics or even destroy it \cite{Gassend-B:2002}. This attack is both costly and impractical since the attacker needs specialized laboratory equipment while the IoT devices could be installed in protected or public areas.

\subsection{Non-invasive attack}

This attack attempts to intercept the communication channel between the device and the server without physical access to the internal components of the IoT device. In this case, the adversary intercept the authentication protocol that forms the challenge-response pairs and develop a machine learning models to predict them (see section \ref{ML attack}). The Arbiter-PUF (section \ref{ArbiterPUF}) is vulnerable to machine learning attack, which works by building a model that learns the correlation of known CRPs and predict the unknown CRPs. The authors in \cite{Yu-Zhuang:2021} introduced a defensive interface that improves the arbiter-PUF and similar design to resist the machine learning attack. Several designs have been proposed to overcome this vulnerability in the arbiter-PUF, such as XOR Arbiter PUF \cite{Suh:2007}\cite{Mursi:2019} and Feed-Forward Arbiter PUF \cite{Gassend-Blaise-Lim:2004} \cite{Lee:2004}\cite{Tajik:2014}. In addition, the authors in \cite{Ruhrmair:2013} evaluated various machine learning models such as, Logistic Regression (LR), Support Vector Machines (SVMs) and Evolution Strategies (ES) against Arbiter PUF. Moreover, the work in \cite{Bahar:2021} demonstrated a a non-invasive attack against SRAM PUF based on chip correlation parameters, identical specification that sharing manufacturing process. The experiment showed that the adversary was able to guess approximately 45\% of the CRPs for SRAM PUF.  

\subsection{Semi-invasive}

This attack tries to access the PUF chip without destruction, so the adversary can apply multiple techniques, such as photonic emission analysis\cite{Tajik:2019}, to physically characterize the arbiter-PUF from the backside. In addition, other techniques, such as laser fault injection and optical contactless probing, have been demonstrated by \cite{Merli:2013} to predict the secret key from the PUF chip. Similarly, the authors in \cite{Bayon:2013} proposed an electromagnetic analysis attack that identifies leakage frequencies of the Ring Oscillator PUF (section \ref{ROPUF}). However, these types of attacks still require specialized laboratory equipment.

\begin{table*}
    \begin{adjustbox}{max width=\textwidth}
    \begin{threeparttable}
    \caption{Summery of machine learning attack models for different PUF architectures}
    \label{tab:table4}
    \begin{tabular}{ c|c|c|c|c|c|c }
    \hline
    \makecell{Ref} & \makecell{PUF \\ Architectures} & \makecell{ML models} & \makecell{Prediction Rate (\%)} & \makecell{Training Time} & \makecell{CRPs} &  \makecell{Bit Length} \\  
    \hline
    \hline
    \cite{Ruhrmair:2013} & Arbiter PUF & LR\tnote{a} & 95 & 0.01s & 640 & 64 \\
    \hline
    \cite{Fang:2018} & Arbiter PUF & \makecell{LR\\ES\tnote{f}\\Naive Bayes\\AdaBoost} & \makecell{79.05\\74.8\\84.30\\83.10} & \makecell{0.0024s\\76.19s\\0.0011s\\0.1167s} & 400 & 64 \\ 
    \hline
    \cite{Ruhrmair:2013} & RO PUF\cite{Suh:2007} & QS\tnote{c} & 99 & --- & 83,941 & 1024 \\ \hline
    \cite{Mursi:2019} & XOR Arbiter PUF & NN\tnote{d} & 50.40\tnote{e} & --- & 24,000 & 64 \\
    \hline
    \cite{Avvaru:2020} & FF XOR PUF & NN\tnote{d} & 50 & --- & 100,000 & 32 \\
    \hline
    \cite{hou:2020} & SRPUF &  \makecell{LR\\ES\tnote{f}\\DL\tnote{g}} & \makecell{49.9\\50.20\\50.28} & \makecell{---\\---\\54 min 30 s} & 500,000 & 64 \\
    \hline
    \cite{Wei:2020} & Transformer PUF & \makecell{LR\\ES} & \makecell{60\\60}& --- & 2000 & 128 \\
    \hline
    \cite{Ali:2021} & MRAM PUF & \makecell{LR\\SVM\\NN} & \makecell{65.12\\64.53\\62.7} & --- & 10,000 & 15 \\
    \hline
    \cite{Alamro:2021} & Multiplexer PUF\tnote{h} & NN & 1.26 & --- & 8.55$\times$10\(^5\) & 32 \\
    \hline
    \cite{Khan:2021} & Memristor PUF & \makecell{LR\\SVM} & \makecell{50.5\tnote{i}\\56.5\tnote{i}} & --- & 5000 & Xbar size 32$\times$2\\  
    \hline
    \end{tabular}

        \begin{tablenotes}
            \item[a] Logistic Regression.
            \item[b] Support Vector Machines.
            \item[c] Quick Sort.
            \item[d] Neural Network.
            \item[e] 8 XOR PUF.
            \item[f] Evolution Strategies.
            \item[g] Deep Learning methods.
            \item[h] 7 stages.
            \item[i] With XORing.
        \end{tablenotes}
    \end{threeparttable}
\end{adjustbox}
\end{table*}

\subsection{Machine learning attack}\label{ML attack}
Machine learning attack is the most popular and effective attacks on PUF. In this attack, the attacker intercepts the communication link between the device and the server without physical access to the internal components of the chip. The attacker then develops a machine learning models to predict the CRPs. Numerous research papers used these attacks against several designs, so designers can evaluate the security strength, weakness and how effectively the attacker can predict the CRPs.

The PUF circuit can be designed to maintain a limited number of challenges in a very short period for each authentication process. However, the challenge-response pair must never be reused to prevent any machine learning attack \cite{uhrmair:2010}. In fact, producing large number of pairs is usually required to prevent such prediction. In this case, the PUF circuit will be large and that will increase computational overhead. To address this issue, the authors in \cite{Babaei:2017} proposed a reconfigurable design that increases the number of pairs without affecting the computational resources of the IoT. 

The aforementioned attack scenarios need to be considered to develop a secure PUF circuit that can resist the machine learning attack and detect any possible invasive attack. Strong PUFs can be used with the secure authentication protocol to satisfy this security requirements. The authors in \cite{Gao:2016} introduced obfuscated challenge response protocol to prevent machine learning attack without conventional cryptography, which consists of PUF chip, random number generator and control block. 
Furthermore, the authors in \cite{Mursi:2019} proposed a training model with Mutli-layer Perceptrons (MLP) as neural network to predict CRPs for XOR Arbiter PUF. The study shows that the accuracy of the prediction depends on the size of the XOR gates (i.e., when the XOR size increase the prediction accuracy percentage decreased). However, the authors believe that prediction rate below 80\% can be considered secure as the authentication process relied on multiple response bits (approximately 64 bits). Similarly, the authors in \cite{Avvaru:2020} demonstrated a neural network attack against feed-forward XOR PUF with 50\% prediction rate for multiple PUF stages. This shows that the proposed feed-forward XOR PUF is considered more secure compared to XOR PUF. Moreover, the authors in \cite{hou:2020} introduced various attack models in their design of configurable LFSR PUF. They performed Logistic Regression (LR), Evolution Strategies (ES) and Neural Network. As a result, the prediction rate is approximately 50\% of the proposed models. In \cite{Wei:2020}, the authors implemented two models attack, LR and ES for 128 stages Transformer PUF with 8 XOR gates. The evaluation achieved 60\% accuracy compared to classical RO PUF with 90\% prediction rate. In addition, the authors in \cite{Sahoo:2015} developed two mathematical attack on previous PUF design, lightweight secure PUF \cite{Majzoobi:2008} and composite PUF \cite{Sahoo:2014}, which consists of multiple different PUF design stages that usually combine strong and weak PUFs. On the other hand, the authors in \cite{Ali:2021} applied SVM, LR and MPL modeling attack on their proposed Spin-transfer torque (STT)-MRAM reconfigurable Arbiter PUF. It was shown that the design reduces the prediction rate to 65.12\% without utilizing XOR gates and 44.34\% with XOR gates. Furthermore, the authors in \cite{Fang:2018} studied the various modeling attack such as, LR, ES, Naive Bayes and AdaBoost on Arbiter PUF with different cases in terms of number of training sets, efficiency of the machine learning algorithms and several number of Arbiter stages. The experiment show that LR and ES performed better for large data sets, while Naive Bayes and AdaBoost applied for small data sets. Additionally, the training time of Naive Bayes faster (0.0007s) compered to other models with the highest prediction rate as shown in table \ref{tab:table4}. Moreover, the multiplexers have been widely used in delay based PUFs for switching between paths. However, the authors in \cite{Alamro:2021} evaluated multiple stages of multiplexer PUF based on Neural Network method to predict the generated responses and showed that the multiplexer PUF is vulnerable to the machine learning attack with high prediction rate. Consequently, the LR and SVM modelling attacks have been analyzed against memristive Xbar PUF in \cite{Khan:2021} under linear and nonlinear (e.g. XOR) architectures. The evaluation of LR based attack show that the XORing nonlinearity drop the prediction rate accuracy from 73\% to 50.5\%, which is near ideal, while exhibiting high resilience against such attack. Table \ref{tab:table4} summarizes the common machine learning attack models and related PUFs with the accuracy of predicting CRPs table.  

\subsection{Side-Channel attack}
The side-channel attack is one of the common and powerful technique used to breach a PUF circuit. This attack relied on the leakage information that can be occupied by the power dissipation during the key generation process. For instance, the adversary can exploit the relation between power dissipation and CRPs on the PUF and measure the correlation between two variables, such as, responses $R$ and correlation coefficient $r$ with the corresponding power dissipation $P$. Several works evaluated the side-channel attack in their proposed design. The authors in \cite{Ali:2021} evaluated the (STT)-MRAM reconfigurable Arbiter PUF by measuring the correlation between power consumption and 800 generated responses. It was shown that there are no correlation associated between the two variables and the design resist side-channel attacks. Furthermore, the authors in \cite{Kroeger:2020} proposed Cross-PUF attacks, which exploit the leakage power of the Latch in Arbiter PUF to train machine learning models, such as SVM, to predict responses. Therefore, the adversary targets one PUF as a references without recording CRPs to breach all PUFs that fabricated from the same Graphic Design System (GDS), which contains a database of circuit layout.

\section{Conclusion and Outlook}\label{Conclusion}
In recent years, advances in PUF architectures provided a solution for solving and enhancing the security of IoT devices. In this paper, we provided an overview of PUF architectures to provide applicable security solutions for IoT environments due to the low computational complexity of PUF circuit design, less energy and improve the quality metrics such as, randomness, uniqueness and reliability. In addition, the PUF based authentication protocols has been discussed and common security concerns and effective attacks against PUF were reviewed.   

More work is still needed to test different properties of PUF to evaluate their security strengths and weakness. The aforementioned techniques provide lightweight authentication for IoT without utilizing the traditional cryptography methods that can increase the power and the resources consumption, still without having to store secret keys in memory. 

More investigations is need to design proper PUF architecture that prevent machine learning attack, which traditional cryptography methods such as Hash function are used to prevent in non PUF solutions. However, confidentiality and integrity based PUF for exchanging information between IoT devices is still not addressed by the PUF community, which makes it largely an open problem. 

In practice,  conventional confidentiality and integrity techniques have been utilized by the PUF authentication protocols, trading off circuit and computational complexity. Therefore, more research is needed to provide suitable and practical encryption and integrity mechanism that can be implemented in the lightweight applications with low energy consumption and more secure.


\section*{Acknowledgments}
This work is partially funded by the G5797 “Developing Physical-Layer Security Schemes for Internet of Things Networks” project under the NATO’s Science for Peace Programme.

\bibliographystyle{unsrt}  
\bibliography{References}  

\begin{thebibliography}{100}

\bibitem{D.wang:2019}
D.~{Wang}, B.~{Bai}, W.~{Zhao}, and Z.~{Han}.
\newblock A survey of optimization approaches for wireless physical layer
  security.
\newblock {\em IEEE Communications Surveys Tutorials}, 21(2):1878--1911, 2019.

\bibitem{J.zhang:2019}
J.~{Zhang}, S.~{Rajendran}, Z.~{Sun}, R.~{Woods}, and L.~{Hanzo}.
\newblock Physical layer security for the internet of things: Authentication
  and key generation.
\newblock {\em IEEE Wireless Communications}, 26(5):92--98, 2019.

\bibitem{mahdi:2020}
Mahdi Shakiba-Herfeh, Arsenia Chorti, and H.~Vince Poor.
\newblock Physical layer security: Authentication, integrity and
  confidentiality, 2020.

\bibitem{Shor-P.W:1994}
P.W. Shor.
\newblock Algorithms for quantum computation: discrete logarithms and
  factoring.
\newblock In {\em Proceedings 35th Annual Symposium on Foundations of Computer
  Science}, pages 124--134, 1994.

\bibitem{N.xie:2021}
N.~{Xie}, Z.~{Li}, and H.~{Tan}.
\newblock A survey of physical-layer authentication in wireless communications.
\newblock {\em IEEE Communications Surveys Tutorials}, 23(1):282--310, 2021.

\bibitem{Schinianakis:2019}
Dimitrios Schinianakis.
\newblock Lightweight security for the internet of things: A soft introduction
  to physical unclonable functions.
\newblock {\em IEEE Potentials}, 38(2):21--28, 2019.

\bibitem{babaei:2019}
Armin Babaei and Gregor Schiele.
\newblock Physical unclonable functions in the internet of things: State of the
  art and open challenges.
\newblock {\em Sensors}, 19(14):3208, Jul 2019.

\bibitem{zhang_qu:2014}
Ji-Liang Zhang, Gang Qu, Yong-Qiang Lv, and Qiang Zhou.
\newblock A survey on silicon pufs and recent advances in ring oscillator pufs.
\newblock {\em Journal of Computer Science and Technology}, 29(4):664–678,
  2014.

\bibitem{Chang:2017}
Chip-Hong Chang, Yue Zheng, and Le~Zhang.
\newblock A retrospective and a look forward: Fifteen years of physical
  unclonable function advancement.
\newblock {\em IEEE Circuits and Systems Magazine}, 17(3):32--62, 2017.

\bibitem{Halak:2016}
Basel Halak, Mark Zwolinski, and M.~Syafiq Mispan.
\newblock Overview of puf-based hardware security solutions for the internet of
  things.
\newblock In {\em 2016 IEEE 59th International Midwest Symposium on Circuits
  and Systems (MWSCAS)}, pages 1--4, 2016.

\bibitem{Herder:2014}
Charles Herder, Meng-Day Yu, Farinaz Koushanfar, and Srinivas Devadas.
\newblock Physical unclonable functions and applications: A tutorial.
\newblock {\em Proceedings of the IEEE}, 102(8):1126--1141, 2014.

\bibitem{Shamsoshoara:2020}
Alireza Shamsoshoara, Ashwija Korenda, Fatemeh Afghah, and Sherali Zeadally.
\newblock A survey on physical unclonable function (puf)-based security
  solutions for internet of things.
\newblock {\em Computer Networks}, 183:107593, 2020.

\bibitem{Alkatheiri:2017}
Mohammed~Saeed Alkatheiri, Yu~Zhuang, Mikhail Korobkov, and Abdur~Rashid Sangi.
\newblock An experimental study of the state-of-the-art pufs implemented on
  fpgas.
\newblock In {\em 2017 IEEE Conference on Dependable and Secure Computing},
  pages 174--180, 2017.

\bibitem{Adames:2016}
Ilia~A. Bautista~Adames, Jayita Das, and Sanjukta Bhanja.
\newblock Survey of emerging technology based physical unclonable funtions.
\newblock In {\em 2016 International Great Lakes Symposium on VLSI (GLSVLSI)},
  pages 317--322, 2016.

\bibitem{Chang:2016}
Chip-Hong Chang and Miodrag Potkonjak.
\newblock {\em Secure System Design and Trustable Computing}.
\newblock Springer International Publishing, 2016.

\bibitem{Suh:2007}
G.~Edward Suh and Srinivas Devadas.
\newblock Physical unclonable functions for device authentication and secret
  key generation.
\newblock In {\em 2007 44th ACM/IEEE Design Automation Conference}, pages
  9--14, 2007.

\bibitem{Tehranipoor:2010}
Mohammad Tehranipoor and Farinaz Koushanfar.
\newblock A survey of hardware trojan taxonomy and detection.
\newblock {\em IEEE Design Test of Computers}, 27(1):10--25, 2010.

\bibitem{armatix:2021}
armatix.
\newblock Armatix ip1 limited edition set.
\newblock \url{http://www.armatix.us/iP1-Limited-Edition.804.0.html?&L=7},
  2021.
\newblock [Online; accessed 2021].

\bibitem{Jeremy-Hsu:2021}
Jeremy Hsu.
\newblock New u.s. military chip self destructs on command.
\newblock
  \url{https://spectrum.ieee.org/tech-talk/computing/hardware/us-militarys-chip-self-destructs-on-command},
  2015.
\newblock [Online; accessed 2021].

\bibitem{Maes:2010}
Roel Maes and Ingrid Verbauwhede.
\newblock Physically unclonable functions: A study on the state of the art and
  future research directions.
\newblock {\em Information Security and Cryptography Towards Hardware-Intrinsic
  Security}, page 3–37, 2010.

\bibitem{uhrmair:2010}
Ulrich~R Uhrmair, Frank Sehnke, Jan~S Olter, Gideon Dror, Srinivas Devadas, and
  J~Urgen Schmidhuber.
\newblock Modeling attacks on physical unclonable functions.
\newblock {\em Proceedings of the 17th ACM conference on Computer and
  communications security - CCS 10}, 2010.

\bibitem{Gassend:2002}
B.~Gassend, D.~Clarke, M.~van Dijk, and S.~Devadas.
\newblock Controlled physical random functions.
\newblock In {\em 18th Annual Computer Security Applications Conference, 2002.
  Proceedings.}, pages 149--160, 2002.

\bibitem{Maiti:2011}
Abhranil Maiti, Vikash Gunreddy, and Patrick Schaumont.
\newblock A systematic method to evaluate and compare the performance of
  physical unclonable functions.
\newblock {\em IACR Cryptology ePrint Archive}, 2011:657, 01 2011.

\bibitem{Kang:2012}
Hyunho Kang, Yohei Hori, and Akashi Satoh.
\newblock Performance evaluation of the first commercial puf-embedded rfid.
\newblock In {\em The 1st IEEE Global Conference on Consumer Electronics 2012},
  pages 5--8, 2012.

\bibitem{Jouini:2011}
Zouha~Cherif Jouini, Jean-Luc Danger, and Lilian Bossuet.
\newblock Performance evaluation of physically unclonable function by delay
  statistics.
\newblock In {\em 2011 IEEE 9th International New Circuits and systems
  conference}, pages 482--485, 2011.

\bibitem{Mahalat:2018}
Mahabub~Hasan Mahalat, Shreya Saha, Anindan Mondal, and Bibhash Sen.
\newblock A puf based light weight protocol for secure wifi authentication of
  iot devices.
\newblock In {\em 2018 8th International Symposium on Embedded Computing and
  System Design (ISED)}, pages 183--187, 2018.

\bibitem{Nakhila:2016}
Omar Nakhila and Cliff Zou.
\newblock User-side wi-fi evil twin attack detection using random wireless
  channel monitoring.
\newblock In {\em MILCOM 2016 - 2016 IEEE Military Communications Conference},
  pages 1243--1248, 2016.

\bibitem{Satamraju:2020}
Krishna~Prasad Satamraju and B.~Malarkodi.
\newblock A puf-based mutual authentication protocol for internet of things.
\newblock In {\em 2020 5th International Conference on Computing, Communication
  and Security (ICCCS)}, pages 1--6, 2020.

\bibitem{Aman:2019}
Muhammad~Naveed Aman, Mohammed~Haroon Basheer, and Biplab Sikdar.
\newblock Data provenance for iot with light weight authentication and privacy
  preservation.
\newblock {\em IEEE Internet of Things Journal}, 6(6):10441--10457, 2019.

\bibitem{Goutsos:2019}
Konstantinos Goutsos and Alex Bystrov.
\newblock Lightweight puf-based continuous authentication protocol.
\newblock In {\em 2019 International Conference on Computing, Electronics
  Communications Engineering (iCCECE)}, pages 229--234, 2019.

\bibitem{Noura:2019}
Hassan~N. Noura, Reem Melki, and Ali Chehab.
\newblock Secure and lightweight mutual multi-factor authentication for iot
  communication systems.
\newblock In {\em 2019 IEEE 90th Vehicular Technology Conference
  (VTC2019-Fall)}, pages 1--7, 2019.

\bibitem{Jiang:2019}
Qi~Jiang, Xin Zhang, Ning Zhang, Youliang Tian, Xindi Ma, and Jianfeng Ma.
\newblock Two-factor authentication protocol using physical unclonable function
  for iov.
\newblock In {\em 2019 IEEE/CIC International Conference on Communications in
  China (ICCC)}, pages 195--200, 2019.

\bibitem{Gassend-B:2002}
Blaise Gassend, Dwaine Clarke, Marten van Dijk, and Srinivas Devadas.
\newblock Silicon physical random functions.
\newblock In {\em Proceedings of the 9th ACM Conference on Computer and
  Communications Security}, CCS '02, page 148–160, New York, NY, USA, 2002.
  Association for Computing Machinery.

\bibitem{Dubrova:2018}
Elena Dubrova.
\newblock A reconfigurable arbiter puf with 4 x 4 switch blocks.
\newblock In {\em 2018 IEEE 48th International Symposium on Multiple-Valued
  Logic (ISMVL)}, pages 31--37, 2018.

\bibitem{Cao:2019}
Yuan Cao, Wenhan Zheng, Xiaojin Zhao, and Chip-Hong Chang.
\newblock An energy-efficient current-starved inverter based strong physical
  unclonable function with enhanced temperature stability.
\newblock {\em IEEE Access}, 7:105287--105297, 2019.

\bibitem{Moradi:2020}
Mona Moradi, Reza~Faghih Mirzaee, and Sha Tao.
\newblock Cmos arbiter physical unclonable function with selecting modules.
\newblock In {\em 2020 20th International Symposium on Computer Architecture
  and Digital Systems (CADS)}, pages 1--6, 2020.

\bibitem{Deng:2020}
Ding Deng, Shen Hou, Zhenyu Wang, and Yang Guo.
\newblock Configurable ring oscillator puf using hybrid logic gates.
\newblock {\em IEEE Access}, 8:161427--161437, 2020.

\bibitem{Qian:2019}
Qian Wang and Gang Qu.
\newblock A silicon puf based entropy pump.
\newblock {\em IEEE Transactions on Dependable and Secure Computing},
  16(3):402--414, 2019.

\bibitem{Zayed:2019}
Amin~A. Zayed, Hanady~H. Issa, and Khaled~A. Shehata.
\newblock Finfet based low power ring oscillator physical unclonable functions.
\newblock In {\em 2019 31st International Conference on Microelectronics
  (ICM)}, pages 227--230, 2019.

\bibitem{D.E.Holcomb:2007}
D.E. Holcomb, W.P. Burleson, and K.~Fu.
\newblock Initial sram state as a fingerprint and source of true random numbers
  for rfid tags.
\newblock {\em Proc. Conf. Radio Frequency Identification Security (RFID
  ’07}, 2007.

\bibitem{Guajardo:2007}
Jorge Guajardo, Sandeep~S. Kumar, Geert-Jan Schrijen, and Pim Tuyls.
\newblock Fpga intrinsic pufs and their use for ip protection.
\newblock In Pascal Paillier and Ingrid Verbauwhede, editors, {\em
  Cryptographic Hardware and Embedded Systems - CHES 2007}, pages 63--80,
  Berlin, Heidelberg, 2007. Springer Berlin Heidelberg.

\bibitem{Sankaran:2018}
Sriram Sankaran, S.~Shivshankar, and K.~Nimmy.
\newblock Lhpuf: Lightweight hybrid puf for enhanced security in internet of
  things.
\newblock In {\em 2018 IEEE International Symposium on Smart Electronic Systems
  (iSES) (Formerly iNiS)}, pages 275--278, 2018.

\bibitem{Yanambaka:2016}
Venkata~P. Yanambaka, Saraju~P. Mohanty, and Elias Kougianos.
\newblock Novel finfet based physical unclonable functions for efficient
  security integration in the iot.
\newblock In {\em 2016 IEEE International Symposium on Nanoelectronic and
  Information Systems (iNIS)}, pages 172--177, 2016.

\bibitem{Pappu:2001}
Pappu~Srinivasa Ravikanth and Stephen~A. Benton.
\newblock Physical one-way functions.
\newblock {\em Science}, 297:2026--2030, 2001.

\bibitem{Pappu:2002}
Ravi Pappu, Ben Recht, Jason Taylor, and Neil~A. Gershenfeld.
\newblock Physical one-way functions.
\newblock {\em Science}, 297:2026 -- 2030, 2002.

\bibitem{michael:2017}
Michael Geis, Karen Gettings, and Michael Vai.
\newblock Optical physical unclonable function.
\newblock {\em 2017 IEEE 60th International Midwest Symposium on Circuits and
  Systems (MWSCAS)}, 2017.

\bibitem{Rhrmair:2013}
Ulrich R{\"u}hrmair, Christian Hilgers, and Sebastian Urban.
\newblock Optical pufs reloaded.
\newblock {\em Eprint.Iacr.Org}, 2013.

\bibitem{Chua:1971}
L.~Chua.
\newblock Memristor-the missing circuit element.
\newblock {\em IEEE Transactions on Circuit Theory}, 18(5):507--519, 1971.

\bibitem{strukov:2008}
Dmitri~B. Strukov, Gregory~S. Snider, Duncan~R. Stewart, and R.~Stanley
  Williams.
\newblock The missing memristor found.
\newblock {\em Nature}, 453(7191):80–83, 2008.

\bibitem{Uddin:2019}
Mesbah Uddin, Md~Sakib Hasan, and Garrett~S. Rose.
\newblock On the theoretical analysis of memristor based true random number
  generator.
\newblock In {\em Proceedings of the 2019 on Great Lakes Symposium on VLSI},
  GLSVLSI '19, page 21–26, New York, NY, USA, 2019. Association for Computing
  Machinery.

\bibitem{Koeberl:2013}
Patrick Koeberl, Unal Kocabaş, and Ahmad-Reza Sadeghi.
\newblock Memristor pufs: A new generation of memory-based physically
  unclonable functions.
\newblock In {\em 2013 Design, Automation Test in Europe Conference Exhibition
  (DATE)}, pages 428--431, 2013.

\bibitem{Rose:2013}
Garrett~S. Rose, Nathan McDonald, Lok-Kwong Yan, Bryant Wysocki, and Karen Xu.
\newblock Foundations of memristor based puf architectures.
\newblock In {\em 2013 IEEE/ACM International Symposium on Nanoscale
  Architectures (NANOARCH)}, pages 52--57, 2013.

\bibitem{RoseA:2013}
Garrett~S. Rose, Nathan McDonald, Lok-Kwong Yan, and Bryant Wysocki.
\newblock A write-time based memristive puf for hardware security applications.
\newblock In {\em 2013 IEEE/ACM International Conference on Computer-Aided
  Design (ICCAD)}, pages 830--833, 2013.

\bibitem{Uddin:2018}
Mesbah Uddin, MD.~Badruddoja Majumder, Karsten Beckmann, Harika Manem,
  Zahiruddin Alamgir, Nathaniel~C. Cady, and Garrett~S. Rose.
\newblock Design considerations for memristive crossbar physical unclonable
  functions.
\newblock {\em J. Emerg. Technol. Comput. Syst.}, 14(1), September 2017.

\bibitem{Muhammad:2021}
Muhammad~Ibrar Khan, Shawkat Ali, Aref Al-Tamimi, Arshad Hassan, Ataul~Aziz
  Ikram, and Amine Bermak.
\newblock A robust architecture of physical unclonable function based on
  memristor crossbar array.
\newblock {\em Microelectronics Journal}, 116:105238, 2021.

\bibitem{Khan:2021}
Muhammad~Ibrar Khan, Shawkat Ali, Ataul~Aziz Ikram, and Amine Bermak.
\newblock Optimization of memristive crossbar array for physical unclonable
  function applications.
\newblock {\em IEEE Access}, 9:84480--84489, 2021.

\bibitem{Boris:2010}
Boris {\v{S}}kori{\'{c}}.
\newblock Quantum readout of physical unclonable functions.
\newblock In Daniel~J. Bernstein and Tanja Lange, editors, {\em Progress in
  Cryptology -- AFRICACRYPT 2010}, pages 369--386, Berlin, Heidelberg, 2010.
  Springer Berlin Heidelberg.

\bibitem{Wootters:1982}
W.~K. Wootters and W.~H. Zurek.
\newblock A single quantum cannot be cloned.
\newblock {\em Nature}, 299(5886):802–803, 1982.

\bibitem{Goorden:2014}
Sebastianus~A. Goorden, Marcel Horstmann, Allard~P. Mosk, Boris \v{S}kori\'{c},
  and Pepijn W.~H. Pinkse.
\newblock Quantum-secure authentication of a physical unclonable key.
\newblock {\em Optica}, 1(6):421--424, Dec 2014.

\bibitem{roberts:2015}
J.~Roberts, I.~E. Bagci, M.~A.~M. Zawawi, J.~Sexton, N.~Hulbert, Y.~J. Noori,
  M.~P. Young, C.~S. Woodhead, M.~Missous, M.~A. Migliorato, and et~al.
\newblock Using quantum confinement to uniquely identify devices.
\newblock {\em Scientific Reports}, 5(1), 2015.

\bibitem{arapinis:2021}
Myrto Arapinis, Mahshid Delavar, Mina Doosti, and Elham Kashefi.
\newblock Quantum physical unclonable functions: Possibilities and
  impossibilities.
\newblock {\em Quantum}, 5:475, 2021.

\bibitem{Phalak:2021}
Koustubh Phalak, Abdullah~Ash Saki, Mahabubul Alam, Rasit~Onur Topaloglu, and
  Swaroop Ghosh.
\newblock Quantum puf for security and trust in quantum computing.
\newblock {\em IEEE Journal on Emerging and Selected Topics in Circuits and
  Systems}, 11(2):333--342, 2021.

\bibitem{Jayita:2014}
Jayita Das, Kevin Scott, Drew Burgett, Srinath Rajaram, and Sanjukta Bhanja.
\newblock A novel geometry based mram puf.
\newblock In {\em 14th IEEE International Conference on Nanotechnology}, pages
  859--863, 2014.

\bibitem{Das:2015}
Jayita Das, Kevin Scott, Srinath Rajaram, Drew Burgett, and Sanjukta Bhanja.
\newblock Mram puf: A novel geometry based magnetic puf with integrated cmos.
\newblock {\em IEEE Transactions on Nanotechnology}, 14(3):436--443, 2015.

\bibitem{Nejat:2020}
Arash Nejat, Frederic Ouattara, Mohammad Mohammadinodoushan, Bertrand Cambou,
  Ken Mackay, and Lionel Torres.
\newblock Practical experiments to evaluate quality metrics of mram-based
  physical unclonable functions.
\newblock {\em IEEE Access}, 8:176042--176049, 2020.

\bibitem{Ali:2021}
Rashid Ali, You Wang, Haoyuan Ma, Zhengyi Hou, Deming Zhang, Erya Deng, and
  Weisheng Zhao.
\newblock A reconfigurable arbiter puf based on stt-mram.
\newblock In {\em 2021 IEEE International Symposium on Circuits and Systems
  (ISCAS)}, pages 1--5, 2021.

\bibitem{Hu:2021}
Yupeng Hu, Linjun Wu, Zhuojun Chen, Yun Huang, Xiaolin Xu, Keqin Li, and
  Jiliang Zhang.
\newblock Stt-mram-based reliable weak puf.
\newblock {\em IEEE Transactions on Computers}, pages 1--1, 2021.

\bibitem{Konigsmark:2014}
S.~T.~Choden Konigsmark, Leslie~K. Hwang, Deming Chen, and Martin D.~F. Wong.
\newblock Cnpuf: A carbon nanotube-based physically unclonable function for
  secure low-energy hardware design.
\newblock In {\em 2014 19th Asia and South Pacific Design Automation Conference
  (ASP-DAC)}, pages 73--78, 2014.

\bibitem{Moradi:2017}
Mona Moradi, Sha Tao, and Reza~Faghih Mirzaee.
\newblock Physical unclonable functions based on carbon nanotube fets.
\newblock In {\em 2017 IEEE 47th International Symposium on Multiple-Valued
  Logic (ISMVL)}, pages 124--129, 2017.

\bibitem{Lee:2019}
Yongwoo Lee, Jinsu Yoon, Hyo-Jin Kim, Geon-Hwi Park, Dae~Hwan Kim, Dong
  Myong~Kim, Min-Ho Kang, and Sung-Jin Choi.
\newblock Carbon nanotube network transistor for a physical unclonable
  functions-based security device.
\newblock In {\em 2019 IEEE 19th International Conference on Nanotechnology
  (IEEE-NANO)}, pages 227--230, 2019.

\bibitem{Srinivasu:2021}
B.~Srinivasu and Anupam Chattopadhyay.
\newblock Cycle puf: A cycle operator based puf in carbon nanotube fet
  technology.
\newblock In {\em 2021 IEEE 21st International Conference on Nanotechnology
  (NANO)}, pages 13--16, 2021.

\bibitem{Suzuki:2010}
Daisuke Suzuki and Koichi Shimizu.
\newblock The glitch puf: A new delay-puf architecture exploiting glitch
  shapes.
\newblock In Stefan Mangard and Fran{\c{c}}ois-Xavier Standaert, editors, {\em
  Cryptographic Hardware and Embedded Systems, CHES 2010}, pages 366--382,
  Berlin, Heidelberg, 2010. Springer Berlin Heidelberg.

\bibitem{Kumar:2008}
Sandeep~S. Kumar, Jorge Guajardo, Roel Maes, Geert-Jan Schrijen, and Pim Tuyls.
\newblock Extended abstract: The butterfly puf protecting ip on every fpga.
\newblock In {\em 2008 IEEE International Workshop on Hardware-Oriented
  Security and Trust}, pages 67--70, 2008.

\bibitem{Su:2007}
Y.~Su, J.~Holleman, and B.~Otis.
\newblock A 1.6pj/bit 96\% stable chip-id generating circuit using process
  variations.
\newblock In {\em 2007 IEEE International Solid-State Circuits Conference.
  Digest of Technical Papers}, pages 406--611, 2007.

\bibitem{Miao:2018}
Jin Miao, Meng Li, Subhendu Roy, Yuzhe Ma, and Bei Yu.
\newblock Sd-puf: Spliced digital physical unclonable function.
\newblock {\em IEEE Transactions on Computer-Aided Design of Integrated
  Circuits and Systems}, 37(5):927--940, 2018.

\bibitem{Tanaka:2018}
Yuki Tanaka, Song Bian, Masayuki Hiromoto, and Takashi Sato.
\newblock Coin flipping puf: A novel puf with improved resistance against
  machine learning attacks.
\newblock {\em IEEE Transactions on Circuits and Systems II: Express Briefs},
  65(5):602--606, 2018.

\bibitem{Gao:2018}
Yansong Gao, Hua Ma, Said~F. Al-Sarawi, Derek Abbott, and Damith~C. Ranasinghe.
\newblock Puf-fsm: A controlled strong puf.
\newblock {\em IEEE Transactions on Computer-Aided Design of Integrated
  Circuits and Systems}, 37(5):1104--1108, 2018.

\bibitem{Zhuang:2020}
Haoyu Zhuang, Xiaodan Xi, Nan Sun, and Michael Orshansky.
\newblock A strong subthreshold current array puf resilient to machine learning
  attacks.
\newblock {\em IEEE Transactions on Circuits and Systems I: Regular Papers},
  67(1):135--144, 2020.

\bibitem{Cao:2021}
Zhen Cao, Shuai Zhang, Jian Zhang, Nuo Xu, Ruofan Li, Zhe Guo, Jijun Yun, Min
  Song, Qiming Zou, Li~Xi, Oukjae Lee, Xiaofei Yang, Xuecheng Zou, Jeongmin
  Hong, and Long You.
\newblock Reconfigurable physical unclonable function based on spin-orbit
  torque induced chiral domain wall motion.
\newblock {\em IEEE Electron Device Letters}, 42(4):597--600, 2021.

\bibitem{Machida:2014}
Takanori Machida, Dai Yamamoto, Mitsugu Iwamoto, and Kazuo Sakiyama.
\newblock A new mode of operation for arbiter puf to improve uniqueness on
  fpga.
\newblock In {\em 2014 Federated Conference on Computer Science and Information
  Systems}, pages 871--878, 2014.

\bibitem{Sushma:2018}
R.~Sushma and N.S. Murty.
\newblock Feedback oriented xored flip-flop based arbiter puf.
\newblock In {\em 2018 International Conference on Electrical, Electronics,
  Communication, Computer, and Optimization Techniques (ICEECCOT)}, pages
  1444--1448, 2018.

\bibitem{Gu:2016}
Chongyan Gu, Yijun Cui, Neil Hanley, and M{\'a}ire O'Neill.
\newblock Novel lightweight ff-apuf design for fpga.
\newblock {\em 2016 29th IEEE International System-on-Chip Conference (SOCC)},
  pages 75--80, 2016.

\bibitem{Sahithi:2018}
Kolasani Sahithi and N.S. Murty.
\newblock Delay based physical unclonable function for hardware security and
  trust.
\newblock In {\em 2018 International Conference on Advances in Computing,
  Communications and Informatics (ICACCI)}, pages 797--803, 2018.

\bibitem{Gu:2019}
Chongyan Gu, Weiqiang Liu, Yijun Cui, Neil Hanley, Maire O'Neill, and Fabrizio
  Lombardi.
\newblock A flip-flop based arbiter physical unclonable function (apuf) design
  with high entropy and uniqueness for fpga implementation.
\newblock {\em IEEE Transactions on Emerging Topics in Computing}, pages 1--1,
  2019.

\bibitem{He:2020}
Zhangqing He, Wanbo Chen, Lingchao Zhang, Gaojun Chi, Qi~Gao, and Lein Harn.
\newblock A highly reliable arbiter puf with improved uniqueness in fpga
  implementation using bit-self-test.
\newblock {\em IEEE Access}, 8:181751--181762, 2020.

\bibitem{Bossuet:2014}
Lilian Bossuet, Xuan~Thuy Ngo, Zouha Cherif, and Viktor Fischer.
\newblock A puf based on a transient effect ring oscillator and insensitive to
  locking phenomenon.
\newblock {\em IEEE Transactions on Emerging Topics in Computing}, 2(1):30--36,
  2014.

\bibitem{Bayon:2013}
Pierre Bayon, Lilian Bossuet, Alain Aubert, and Viktor Fischer.
\newblock Electromagnetic analysis on ring oscillator-based true random number
  generators.
\newblock In {\em 2013 IEEE International Symposium on Circuits and Systems
  (ISCAS)}, pages 1954--1957, 2013.

\bibitem{Yan:2017}
Wei Yan, Chenglu Jin, Fatemeh Tehranipoor, and John~A. Chandy.
\newblock Phase calibrated ring oscillator puf design and implementation on
  fpgas.
\newblock In {\em 2017 27th International Conference on Field Programmable
  Logic and Applications (FPL)}, pages 1--8, 2017.

\bibitem{Garcia-Bosque:2020}
Miguel Garcia-Bosque, Guillermo Díez-SEñorans, Carlos Sánchez-Azqueta, and
  Santiago Celma.
\newblock Proposal and analysis of a novel class of pufs based on galois ring
  oscillators.
\newblock {\em IEEE Access}, 8:157830--157839, 2020.

\bibitem{Li:2020}
Jin Li, Lei Li, Ji~Yang, Yuanhang He, Wanting Zhou, and Shiwei Yuan.
\newblock An efficient and stable composed entropy extraction method for
  fpga-based ro puf.
\newblock {\em IEICE Electronics Express}, 17(24):20200350--20200350, 2020.

\bibitem{Aguirre:2020}
Abby Aguirre, Michael Hall, Timothy Lim, Jonathan Trinh, Wei Yan, and Fatemeh
  Tehranipoor.
\newblock A systematic approach for internal entropy boosting in delay-based ro
  puf on an fpga.
\newblock In {\em 2020 IEEE 63rd International Midwest Symposium on Circuits
  and Systems (MWSCAS)}, pages 623--626, 2020.

\bibitem{Yao:2021}
Liang Yao, Huaguo Liang, Zhengfeng Huang, Cuiyun Jiang, Maoxiang Yi, and
  Yingchun Lu.
\newblock A lightweight configurable xor ro-puf design based on xilinx fpga.
\newblock {\em 2021 IEEE 4th International Conference on Electronics Technology
  (ICET)}, pages 83--88, 2021.

\bibitem{hou:2020}
Shen Hou, Ding Deng, Zhenyu Wang, Jiahe Shi, Shaoqing Li, and Yang Guo.
\newblock A dynamically configurable lfsr-based puf design against machine
  learning attacks.
\newblock {\em CCF Transactions on High Performance Computing}, 3(1):31–56,
  2020.

\bibitem{Hou:2019}
Shen Hou, Yang Guo, and Shaoqing Li.
\newblock A lightweight lfsr-based strong physical unclonable function design
  on fpga (january 2019).
\newblock {\em IEEE Access}, PP:1--1, 05 2019.

\bibitem{Anderson:2010}
Jason~H. Anderson.
\newblock A puf design for secure fpga-based embedded systems.
\newblock In {\em 2010 15th Asia and South Pacific Design Automation Conference
  (ASP-DAC)}, pages 1--6, 2010.

\bibitem{Amsaad:2018}
Fathi Amsaad, Ahmed Sherif, Amer Dawoud, Mohammed Niamat, and Selck Kose.
\newblock A novel fpga-based lfsr puf design for iot and smart applications.
\newblock In {\em NAECON 2018 - IEEE National Aerospace and Electronics
  Conference}, pages 99--104, 2018.

\bibitem{Zhou:2020}
Ting Zhou, Yuxin Ji, Mingyi Chen, and Yongfu Li.
\newblock Pl-mro puf: High speed pseudo-lfsr puf based on multiple ring
  oscillators.
\newblock In {\em 2020 IEEE International Symposium on Circuits and Systems
  (ISCAS)}, pages 1--5, 2020.

\bibitem{Ardakani:2016}
Amir Ardakani and Shahriar Baradaran~Shokouhi.
\newblock A secure and area-efficient fpga-based sr-latch puf.
\newblock In {\em 2016 8th International Symposium on Telecommunications
  (IST)}, pages 94--99, 2016.

\bibitem{Wang:2018}
Wenxuan Wang, Aijiao Cui, Gang Qu, and Huawei Li.
\newblock A low-overhead puf based on parallel scan design.
\newblock In {\em 2018 23rd Asia and South Pacific Design Automation Conference
  (ASP-DAC)}, pages 715--720, 2018.

\bibitem{Shen:2019}
Shen Hou, Yang Guo, Shaoqing Li, Ding Deng, and Yan Lei.
\newblock A lightweight and secure-enhanced strong puf design on fpga.
\newblock {\em IEICE Electronics Express}, 16(24):20190695--20190695, 2019.

\bibitem{Cui:2019}
Yijun Cui, Chenghua Wang, Yunpeng Chen, Ziwei Wei, Mengxian Chen, and Weiqiang
  Liu.
\newblock Dynamic reconfigurable pufs based on fpga.
\newblock In {\em 2019 IEEE International Workshop on Signal Processing Systems
  (SiPS)}, pages 79--84, 2019.

\bibitem{Wei:2020}
Ziwei Wei, Yijun Cui, Yunpeng Chen, Chenghua Wang, Chongyan Gu, and Weiqiang
  Liu.
\newblock Transformer puf : A highly flexible configurable ro puf based on
  fpga.
\newblock In {\em 2020 IEEE Workshop on Signal Processing Systems (SiPS)},
  pages 1--6, 2020.

\bibitem{kalya:2020}
Manasa kalya and Sathish Kumar.
\newblock Low complexity ldpc error correction code for modified anderson puf
  to improve its uniformity.
\newblock In {\em 2020 International Conference on Smart Electronics and
  Communication (ICOSEC)}, pages 997--1002, 2020.

\bibitem{Lotfy:2021}
Armin Lotfy, Masoud Kaveh, Diego Martín, and Mohammad~Reza Mosavi.
\newblock An efficient design of anderson puf by utilization of the xilinx
  primitives in the slicem.
\newblock {\em IEEE Access}, 9:23025--23034, 2021.

\bibitem{Lee:2004}
J.W. Lee, Daihyun Lim, B.~Gassend, G.E. Suh, M.~van Dijk, and S.~Devadas.
\newblock A technique to build a secret key in integrated circuits for
  identification and authentication applications.
\newblock In {\em 2004 Symposium on VLSI Circuits. Digest of Technical Papers
  (IEEE Cat. No.04CH37525)}, pages 176--179, 2004.

\bibitem{Daihyun:2005}
Daihyun Lim, J.W. Lee, B.~Gassend, G.E. Suh, M.~van Dijk, and S.~Devadas.
\newblock Extracting secret keys from integrated circuits.
\newblock {\em IEEE Transactions on Very Large Scale Integration (VLSI)
  Systems}, 13(10):1200--1205, 2005.

\bibitem{crocus:2021}
Crocus technology.
\newblock \url{https://crocus-technology.com/}, Accessed 2021.

\bibitem{Yu-Zhuang:2021}
Yu~Zhuang, Khalid~T. Mursi, and Li~Gaoxiang.
\newblock A challenge obfuscating interface for arbiter puf variants against
  machine learning attacks, 2021.

\bibitem{Mursi:2019}
Khalid~T. Mursi, Yu~Zhuang, Mohammed~Saeed Alkatheiri, and Ahmad~O. Aseeri.
\newblock Extensive examination of xor arbiter pufs as security primitives for
  resource-constrained iot devices.
\newblock In {\em 2019 17th International Conference on Privacy, Security and
  Trust (PST)}, pages 1--9, 2019.

\bibitem{Gassend-Blaise-Lim:2004}
Blaise Gassend, Daihyun Lim, Dwaine Clarke, Marten~Van Dijk, and Srinivas
  Devadas.
\newblock Identification and authentication of integrated circuits.
\newblock {\em Concurrency and Computation: Practice and Experience},
  16(11):1077–1098, 2004.

\bibitem{Tajik:2014}
Shahin Tajik, Enrico Dietz, Sven Frohmann, Jean-Pierre Seifert, Dmitry
  Nedospasov, Clemens Helfmeier, Christian Boit, and Helmar Dittrich.
\newblock Physical characterization of arbiter pufs.
\newblock {\em Advanced Information Systems Engineering Lecture Notes in
  Computer Science}, page 493–509, 2014.

\bibitem{Ruhrmair:2013}
Ulrich Rührmair, Jan Sölter, Frank Sehnke, Xiaolin Xu, Ahmed Mahmoud, Vera
  Stoyanova, Gideon Dror, Jürgen Schmidhuber, Wayne Burleson, and Srinivas
  Devadas.
\newblock Puf modeling attacks on simulated and silicon data.
\newblock {\em IEEE Transactions on Information Forensics and Security},
  8(11):1876--1891, 2013.

\bibitem{Bahar:2021}
B.~M.~S. Bahar~Talukder, Farah Ferdaus, and Md~Tauhidur Rahman.
\newblock Memory-based pufs are vulnerable as well: A non-invasive attack
  against sram pufs.
\newblock {\em IEEE Transactions on Information Forensics and Security},
  16:4035--4049, 2021.

\bibitem{Tajik:2019}
Shahin Tajik.
\newblock On the physical security of physically unclonable functions.
\newblock {\em T-Labs Series in Telecommunication Services}, 2019.

\bibitem{Merli:2013}
Dominik Merli, Johann Heyszl, Benedikt Heinz, Dieter Schuster, Frederic Stumpf,
  and Georg Sigl.
\newblock Localized electromagnetic analysis of ro pufs.
\newblock {\em 2013 IEEE International Symposium on Hardware-Oriented Security
  and Trust (HOST)}, 2013.

\bibitem{Fang:2018}
Yue Fang, Chenghua Wang, Qingqing Ma, Chongyan Gu, Maire O’Neill, and
  Weiqiang Liu.
\newblock Attacking arbiter pufs using various modeling attack algorithms: A
  comparative study.
\newblock In {\em 2018 IEEE Asia Pacific Conference on Circuits and Systems
  (APCCAS)}, pages 394--397, 2018.

\bibitem{Avvaru:2020}
S.~V.~Sandeep Avvaru, Ziqing Zeng, and Keshab~K. Parhi.
\newblock Homogeneous and heterogeneous feed-forward xor physical unclonable
  functions.
\newblock {\em IEEE Transactions on Information Forensics and Security},
  15:2485--2498, 2020.

\bibitem{Alamro:2021}
Meznah~A. Alamro and Khalid~T. Mursi.
\newblock Machine learning attack on a multiplexer puf variant using silicon
  data: a case study on rmpufs.
\newblock In {\em 2021 IEEE 6th International Conference on Computer and
  Communication Systems (ICCCS)}, pages 1017--1022, 2021.

\bibitem{Babaei:2017}
Armin Babaei and Gregor Schiele.
\newblock Spatial reconfigurable physical unclonable functions for the internet
  of things.
\newblock {\em Security, Privacy, and Anonymity in Computation, Communication,
  and Storage Lecture Notes in Computer Science}, page 312–321, 2017.

\bibitem{Gao:2016}
Yansong Gao, Gefei Li, Hua Ma, Said~F. Al-Sarawi, Omid Kavehei, Derek Abbott,
  and Damith~C. Ranasinghe.
\newblock Obfuscated challenge-response: A secure lightweight authentication
  mechanism for puf-based pervasive devices.
\newblock {\em 2016 IEEE International Conference on Pervasive Computing and
  Communication Workshops (PerCom Workshops)}, 2016.

\bibitem{Sahoo:2015}
Durga~Prasad Sahoo, Phuong~Ha Nguyen, Debdeep Mukhopadhyay, and Rajat~Subhra
  Chakraborty.
\newblock A case of lightweight puf constructions: Cryptanalysis and machine
  learning attacks.
\newblock {\em IEEE Transactions on Computer-Aided Design of Integrated
  Circuits and Systems}, 34(8):1334--1343, 2015.

\bibitem{Majzoobi:2008}
Mehrdad Majzoobi, Farinaz Koushanfar, and Miodrag Potkonjak.
\newblock Lightweight secure pufs.
\newblock In {\em 2008 IEEE/ACM International Conference on Computer-Aided
  Design}, pages 670--673, 2008.

\bibitem{Sahoo:2014}
Durga~Prasad Sahoo, Sayandeep Saha, Debdeep Mukhopadhyay, Rajat~Subhra
  Chakraborty, and Hitesh Kapoor.
\newblock Composite puf: A new design paradigm for physically unclonable
  functions on fpga.
\newblock In {\em 2014 IEEE International Symposium on Hardware-Oriented
  Security and Trust (HOST)}, pages 50--55, 2014.

\bibitem{Kroeger:2020}
Trevor Kroeger, Wei Cheng, Sylvain Guilley, Jean-Luc Danger, and Naghmeh
  Karimi.
\newblock Cross-puf attacks on arbiter-pufs through their power side-channel.
\newblock In {\em 2020 IEEE International Test Conference (ITC)}, pages 1--5,
  2020.

\end{thebibliography}

\end{document}